\documentclass[aip,pop,reprint,floats,twocolumn,showpacs]{revtex4-1}
\usepackage[utf8]{inputenc}
\usepackage[english]{babel}
\usepackage{amssymb}
\usepackage{amsmath}
\usepackage[pdftex]{graphicx}
\usepackage{cmap}
\usepackage[pdftex]{hyperref}

 \def\bc{\begin{center}}          \def\ec{\end{center}}
 \allowdisplaybreaks

\begin{document}

\title{Highly efficient electromagnetic emission during 100 keV electron beam relaxation in a thin magnetized plasma }
 \author{V. V. Annenkov}
  \affiliation{Budker Institute of Nuclear Physics SB RAS, 630090, Novosibirsk, Russia}
 \affiliation{Novosibirsk State University, 630090, Novosibirsk, Russia}
  \author{I. V. Timofeev}
 \affiliation{Budker Institute of Nuclear Physics SB RAS, 630090, Novosibirsk, Russia}
 \affiliation{Novosibirsk State University, 630090, Novosibirsk, Russia}
  \author{E. P. Volchok}
 \affiliation{Budker Institute of Nuclear Physics SB RAS, 630090, Novosibirsk, Russia}
 \affiliation{Novosibirsk State University, 630090, Novosibirsk, Russia}

 \begin{abstract}

In this paper, electromagnetic emissions produced by a beam-plasma system are investigated using particle-in-cell simulations for the particular case when 
the typical transverse size of both 100 keV electron beam and produced plasma channel is comparable to the radiation wavelength. The interest in this regime of beam-plasma interaction is associated with highly efficient generation of electromagnetic waves near the plasma frequency harmonics that has been recently observed in laboratory experiments on the GOL-3 mirror trap. It has been found that the radiation power only from the vicinity of the doubled plasma frequency in these experiments can reach 1\% of the total beam power. Subsequent theoretical and simulation studies have shown that the most likely candidate for explaining such efficient generation of electromagnetic radiation is the mechanism of a beam-driven plasma antenna based on the conversion of the most unstable plasma oscillations on a longitudinal density modulation of plasma ions. In this paper, we investigate how effectively this mechanism can work in a real experiment at the GOL-3 facility, when a thin sub-relativistic electron beam gets a large angular spread due to compression by a magnetic field, and the gas into which it is injected has macroscopic density gradients.

 \end{abstract}
 \maketitle

\section{Introduction}

Generation of electromagnetic (EM) radiation  in a plasma by an electron beam has long attracted attention of researchers in relation to both astrophysical phenomena \cite{Gurnett1976,Goldman1980,Whelan1985} and  laboratory experiments \cite{Benford1980,Whelan1981,Pritchett1983,Cheung1982}. Currently, interest in this problem is associated with the study of type II and type III solar radio bursts \cite{Che,Li2013,Robinson1998,Thejappa1998,Thurgood2015} and zebra-like fine spectral structures of solar radiation \cite{Chernov2010,Kuznetsov2013}, as well as with the search for promising schemes for generating high-power sub-THz and THz radiation in mirror traps  \cite{Postupaev2011,Timofeev2012c,Arzhannikov2014,Arzhannikov2016}. Since open magnetic systems allow to inject multigigawatt electron beams, the conversion of even a small fraction of their power into radiation power (1-5\%) would open up the possibility of generating THz pulses at the record gigawatt level.

Recent experiments on the steady-state injection of a long sub-relativistic electron beam with the energy 100 keV and electric current 20-100 A into a magnetized plasma at the multi-mirror trap GOL-3 (BINP SB RAS) \cite{Burdakov2013,Postupaev2013,Ivanov2015} have demonstrated high emission efficiency which seems unusual for a turbulent beam-plasma system. According to authors' estimates, only in the vicinity of the doubled plasma frequency $2\omega_p$, the power of EM emission has reached 1\% of the beam power.  A distinctive feature of these experiments from a conventional regime with kiloampere beams was the low plasma density ($\sim 10^{13}\ \mbox{cm}^{-3}$) in which diameters of both the injected beam and ionized plasma channel ($<1$ cm) turn out to be comparable to wavelengths of plasma oscillations and radiated EM waves.

To explain the efficient generation of radiation in such a thin system, a model of the so called beam-driven plasma antenna has been proposed  \cite{Timofeev2015,Annenkov2016a,Timofeev2016a}. It has been shown that  emission of EM waves from a thin plasma channel becomes possible if the plasma density is longitudinally modulated within a limited range of wavenumbers. Scattering of the most unstable beam-driven plasma wave on this density modulation results in the formation of the superluminal satellite. Due to the beam-induced frequency shift, this satellite is able to transfer its energy to vacuum EM waves with high efficiency only in a relatively thin plasma comparable in size to the skin-depth. An analytical theory and PIC simulations for the beam relaxation in a premodulated plasma have shown that the efficiency of the fundamental electromagnetic emission via this mechanism can reach 5\%-10\%. PIC simulations of the continuous injection of a relativistic electron beam into an initially homogeneous plasma channel \cite{Annenkov2016b} have demonstrated  that the quasi-regular periodic density profile necessary for the antenna emission can arise in a plasma self-consistently due to the modulational instability of the dominant beam-driven mode.

However, it remained not obvious that the same mechanism is able to provide high efficiency of electromagnetic emission at the second harmonic of the plasma frequency, since such an emission should appear as a result of nonlinear interaction of a primary beam wave with a satellite reflected from a density modulation. Generalization of the   plasma antenna theory  to this case has shown that, besides the product of amplitudes of a primary $E_0$ and reflected  $E_1\propto \delta n E_0$ waves, the nonlinear radiating current $j_{2\omega_p} \propto  E_0 E_1/\eta_b$ also contains division by the plasma permittivity $\eta_b=1-\omega_p^2/\omega_b^2$ which is a small parameter at the frequency of beam pumping $\omega_b=\omega_p - \delta$. Due to this fact, the fundamental and second harmonic EM emissions  in the antenna mechanism can be produced with comparable efficiencies. The possibility of achieving high efficiency (2\%-3\%) of beam-to-radiation power conversion near the second harmonic of the plasma frequency has been confirmed in recent PIC simulations \cite{Annenkov2018a}.

The aim of this paper is not only to confirm the principal possibility of efficient (at the 1\% level) generation of $2\omega_p$-radiation  for the typical parameters of the GOL-3 experiments \cite{Burdakov2013} by modeling this phenomenon from first principles, but also to build a general scenario of collective beam relaxation in a plasma to explain qualitatively the results of radiometric measurements at the GOL-3 facility.
The first attempt to simulate injection of a thin 100 keV electron beam in a homogenious plasma \cite{Annenkov2016c} with parameters and scales of laboratory experiments has shown that the region of intense beam relaxation and efficient EM emission at the plasma frequency harmonics is localized near the injector and its size does not exceed a few centimeters. This result contradicts the experimental measurements \cite{Burdakov2013} according to which intense radiation has been observed at a distance of 84 cm from the point of beam entry into the facility. In the present paper, as the reasons  influencing on the local disruption of the beam-plasma instability and shifting the radiation region away from the injector, we will consider (i) a large angular spread of the beam, which is acquired during the beam transport to a strong magnetic field of the plasma trap, (ii) a regular longitudinal gas density inhomogeneity associated with the features of the gas inlet into the facility, and (iii) a high level of small-scale density fluctuations arising due to development of turbulence.

The possibility of reducing the rate of electron beam relaxation in a plasma with an inhomogeneous density profile has been known for a long time \cite{Ryutov1969,Nishikawa1976}. At present, effects of inhomogeneous plasma density are actively considered in astrophysical problems \cite{Krafft2014,Pechhacker2014,Thurgood2016,Schmitz2013,Pechhacker2012} and  laboratory experiments on the generation of THz radiation \cite{Sheng2005} due to the possibility of plasma oscillations to transform into EM waves through the linear mode conversion. The effect of plasma density gradients is also of interest for experiments on the plasma wakefield acceleration, in particular, for the study of the self-modulation instability of a proton beam \cite{Petrenko2016} and electrons injection into an accelerating wake \cite{Faure2010}.

Section \ref{sec:PIC} of this paper contains a brief description of the numerical model and algorithm of creating a realistic momentum particle distribution formed due to compression of an electron beam by a strong magnetic field. Further, we simulate generation of EM radiation in a magnetized plasma column 6 mm wide by a continuously injected sub-relativistic electron beam with the parameters of the GOL-3 experiment (Sec. \ref{sec:Uni}). Then, the possibility to disrupt the beam-plasma instability  by both small-scale plasma density perturbations arising due to development of turbulence and large-scale regular density gradients caused by the initial nonuniform gas distribution  in the facility is investigated in Section \ref{sec:Grad}. The obtained results are further used to build a scenario of beam relaxation in the GOL-3 experiments (Sec. \ref{Experiment}).

\section{Numerical model}\label{sec:PIC}
For modeling dynamics of a beam-plasma system, we use our own parallel 2D3V particle-in-cell code for Cartesian geometry implemented on Nvidia GPGPU\cite{Lindholm2008}. Maxwell equations for EM fields are solved by the standard FDTD scheme of Yee\cite{Yee1966}. To calculate dynamics and currents of finite-size macro-particles with the parabolic form-factor, we use the algorithm of Boris\cite{Boris1970} and the charge conserving density decomposition scheme of Esirkepov\cite{Esirkepov2001}. The grid and time steps are set to  $h=0.04\, c/\omega_p$ and $\tau=0.02\,\omega_p^{-1}$, where $\omega_p=\sqrt{4\pi e^2 n_0/m_e}$ is the plasma frequency, $n_0$ is the unperturbed plasma density, $c$ is the speed of light, $e$ and $m_e$ are the charge and mass of an electron. At the initial time step, ion-electron pairs of hydrogen plasma are placed in the same spatial positions and evenly spaced in the simulation region, while all EM fields  (except the guiding magnetic field $B_x$) are equal to zero.

\subsection{Simulation layout}

\begin{figure}[h!]
	\includegraphics[width=\linewidth]{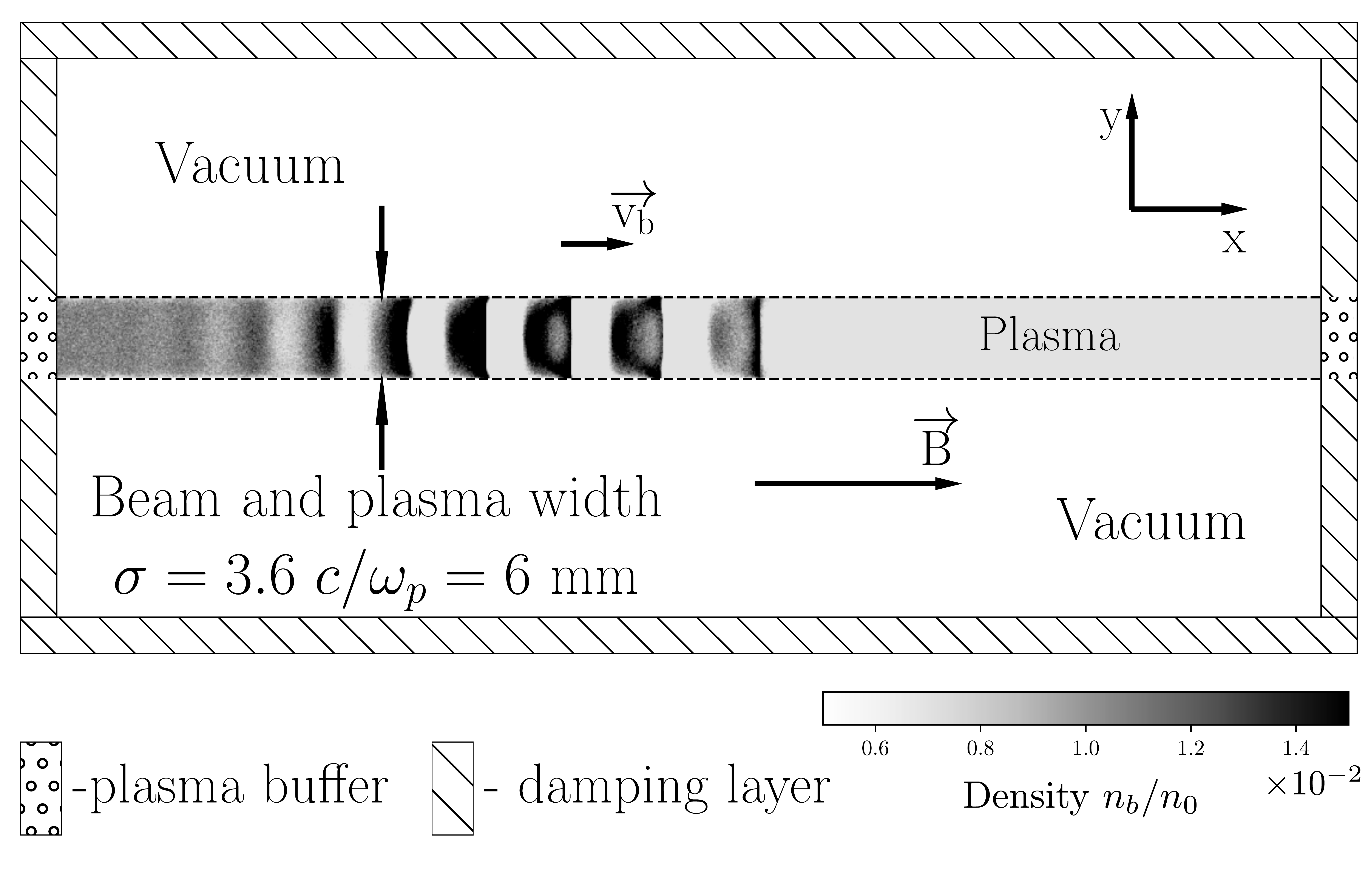} \caption{Simulation layout.}\label{pic:SimBox}
\end{figure}

The simulation layout and beam density in the moment soon after the start of injection are presented in Fig. \ref{pic:SimBox}. 
Damping layers for EM radiation are placed near all borders of the simulation box. A plasma column is located in the middle of the system and separated in the transverse direction from the boundaries by vacuum layers. In the longitudinal direction, plasma is limited by plasma buffers in which the particle distribution similar to the ones inside the boundary plasma cells is maintained. New beam particles are created in these plasma buffers at each time step and then injected into the plasma. If any plasma particle  goes beyond the edge of plasma column in the longitudinal  direction, it is removed from the simulation. A more detailed description of the open boundary conditions and absorbing layers implemented in this code can be found in\cite{Annenkov2018}.

In the laboratory experiments at the GOL-3 facility, plasma has been created via gas ionization by an electron beam itself, therefore their transverse sizes coincide with each other. In presented simulations, we set this size equal to $\sigma=3.6$ $c/\omega_p$, which corresponds to $6$ mm for the plasma density $n_0=10^{13}$ cm$^{-3}$. Vacuum regions have a size $12\,c/\omega_p\approx2.71$ cm. The total system length is $L_x=60\, c/\omega_p\approx10$ cm.  

Plasma electrons have the Maxwellian momentum distribution $f_e\propto\exp\left(-\mathbf{p}^2/(2\Delta p^2_e)\right)$ with the initial temperature $T_e=\Delta p_e^2/(2m_e)=50\,eV$. Plasma ions have the real mass $m_i=1836m_e$ and are initially cold. We use 196 macroparticles in a cell for each particle sort.

The beam density is $n_b=0.01n_0$. For an axially symmetric beam with the diameter $6$ mm and energy 100 keV, this density corresponds to the electric current $I\approx75$ A. Since we are not interested in transient processes associated with a smooth growth of beam current, we inject a beam with a sharp front that creates a seed for the development of the two-stream instability and contributes to a more rapid transition to a nonlinear saturation state.

\subsection{Electron beam distribution}

Since the details of the beam momentum distribution significantly affect both the growth rate of unstable oscillations and their level of nonlinear saturation, we will try to reproduce these details in numerical simulations as realistically as possible. In the real experiment, a beam is created  by a diode with a plasma emitter in the region with the low external magnetic field ($0.01$ T). After acceleration to the required energy, the beam electrons remain almost monoenergetic, but acquire an angular distribution with a characteristic scatter $\Delta\theta=0.015$\cite{Burdakov2013,Astrelin2016}. Further transportation of the beam to the region of a strong magnetic field leads to  compression of its transverse size in hundreds of times and is accompanied by a significant increase in the angular spread.

\begin{figure}[htb]
	\includegraphics[width=\linewidth]{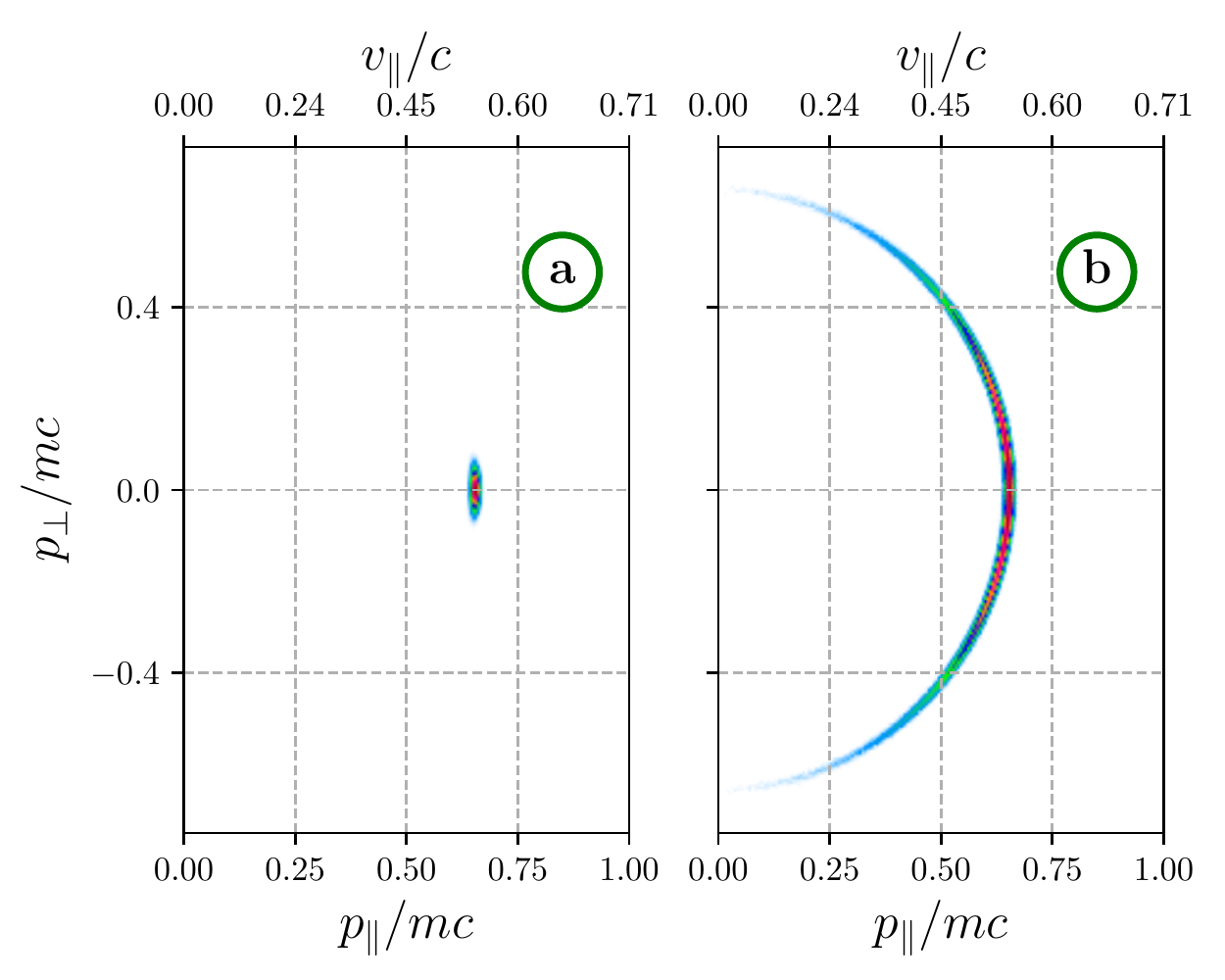} \caption{Particle distribution of the beam on the diode ($B=0.01$ T)  (a) and after transportation to the strong magnetic field of the trap $B=1$ T (b).}\label{pic:Distr}
\end{figure}

In our simulations, such a beam distribution is created in the following way:
\begin{enumerate}
	\item  the momentum of each particle is set in accordance with the shifted Maxwell distribution $f_b(\textbf{p})\propto\exp\left(-(\textbf{p}-\textbf{p}_0)^2/(2\Delta p_b^2)\right)$ with the temperature  $T_b=\Delta p_b^2/(2m_e)=13$ eV and directed momentum $\textbf{p}_0=(p_0,0,0)$ corresponding to the energy of 100 keV ($p_0=0.655 m_e c$);
	\item then this momentum is distributed between the longitudinal and transverse components based on the distribution $f_b(\theta)\propto\exp\left(-\theta^2/(2\Delta \theta^2)\right)$, where $\Delta\theta=0.015$ is the initial angle spread ($p_{0\perp}=p_0\sin\theta$, $p_{0\parallel}=p_0\cos\theta$)  (fig. \ref{pic:Distr} a);	
	\item transporting a beam from a weak ($B_0 = 0.01$ T) to a strong magnetic field $B$ implies the conservation of the magnetic moment and energy of the particles, therefore the final values of their transverse and longitudinal momenta are defined as	$p_{1\perp}=p_{0\perp}\sqrt{B/B_0}$, $p_{1\parallel}=\sqrt{p_0^2-p_{1\perp}^{2}}$ (fig. \ref{pic:Distr} b) (in the case of particle reflection, $p_0^2-p_{1\perp}^2<0$, its momentum is calculated anew in order to save the specified beam current value);
	\item in the Cartesian coordinates we use, the particle momentum vector is represented as three components: $p_x=p_{1\parallel}$, $p_y=p_{1\perp}\cos\left(\pi R\right)$ and $p_z=p_{1\perp}\sin\left(\pi R\right)$, where $R$ is a random number from the uniform distribution from $0$ to $1$.
\end{enumerate}

\section{Injection into a homogeneous plasma}\label{sec:Uni}

We have recently shown \cite{Timofeev2015,Annenkov2016a,Timofeev2016a} that the high efficiency of EM radiation generation in a thin beam-plasma system is possible in the presence of a longitudinal modulation of the plasma density allowing the system to radiate like a dipole antenna. The principal points of this theory are given below.

\subsection{The mechanism of a beam-plasma antenna}

Let us consider a plane plasma layer with the thickness $2l$ and density  $n(x)=n_0+\delta n\sin(qx)$. An electron beam propagating  along the axis $x$ drives longitudinal plasma waves with the frequency $\omega_b$ and the wavenumber $k_\parallel=\omega_b/v_b$. Scattering of these oscillations  on the density  perturbation with the wavenumber $q$ results in excitation of plasma oscillations $(\omega_b,k_\parallel-q)$. Such oscillations can have  superluminal phase velocities and are able to get into the resonance with vacuum EM waves traveling at an angle to the plasma layer. This  resonance is possible only if the period of  density modulation  is not much different from the wavelength of the beam-driven mode 
\begin{equation}\label{Anwp}
1-v_b<q/k_\parallel<1+v_b.
\end{equation} 
The ratio between $q$ and  $k_\parallel$ within this range determines uniquely the angle of radiation 
\begin{equation}
\theta=\arctan\left(\dfrac{\sqrt{v_b^2-(1-q/k_\parallel)^2}}{1-q/k_\parallel}\right).
\end{equation} 

In the particular case $q=k_\parallel$, EM radiation escapes from the plasma in the strictly transverse direction. Since the frequency of the dominant mode in the hydrodynamic regime of the two-stream instability is less than the plasma frequency,
\begin{equation}\label{fr}
\dfrac{\omega_b}{\omega_p}=1-\dfrac{n_b^{1/3}}{2^{4/3}\gamma_b},
\end{equation}
the generated radiation is able to efficiently interact with plasma currents only at the skin depth ($\gamma_b$ is  the relativistic factor of the electrons beam). The radiation power near the plasma frequency  is proportional to the square of the plasma oscillations amplitude \cite{Annenkov2016b}: $P_{\omega_b}{\propto} \delta n^2 E_0^2(x)$. We will determine this amplitude from simulation fields as
$$E_0(x,t)=\left[\dfrac{\omega_b}{\pi}\int\limits_{-\omega_b/\pi}^{\omega_b/\pi}dt'E_x^2(x,t+t')\right]^{1/2}.$$

Besides the fundamental radiation near the plasma frequency, such a system can efficiently produce the second harmonic emission through the nonlinear interaction of the dominant beam-driven mode $(\omega_b,k_\parallel)$ with its long-wavelength satellite $(\omega_b,k_\parallel-q)$ arising due to scattering on the periodic modulation of ion density:
\begin{equation}
(\omega_b,k_\parallel)+(\omega_b,k_\parallel-q)\rightarrow(2\omega_b,2k_\parallel-q).
\end{equation}
Resulting oscillations of electric current $j{\propto}\exp\left(i(2k_\parallel-q)x-i2\omega_bt\right)$ can be a source of EM emission only in the limited range of  $q$ values ($1-v_b<q/(2k_\parallel)<1+v_b$). Thus, for the modulation period corresponding to the range
\begin{equation}\label{An2wp}
1+v_b<q/k_\parallel<2(1+v_b),
\end{equation}
the main part of EM power should be radiated near the second harmonic of the plasma frequency.
The radiation angle is determined by the following expression
\begin{equation}
\theta=\arctan\left(\dfrac{\sqrt{v_b^2-(1-q/(2k_\parallel))^2}}{1-q/(2k_\parallel)}\right).
\end{equation} 
In this case, the radiation power is proportional to the fourth power of the amplitude of 
plasma oscillations \cite{Annenkov2018a}: $P_{2\omega_b}\propto \delta n^2 E_0^4(x)/\eta_b^2$. Despite of the nonlinear character of this process, the power of $2\omega_p$-radiation can be comparable to the power $P_{\omega_b}$ of linear conversion because of a small difference between the pump and plasma frequency ($\eta_b=1-\omega_p^2/\omega_b^2\ll 1$).

\begin{figure}[htb]
	\includegraphics[width=\linewidth]{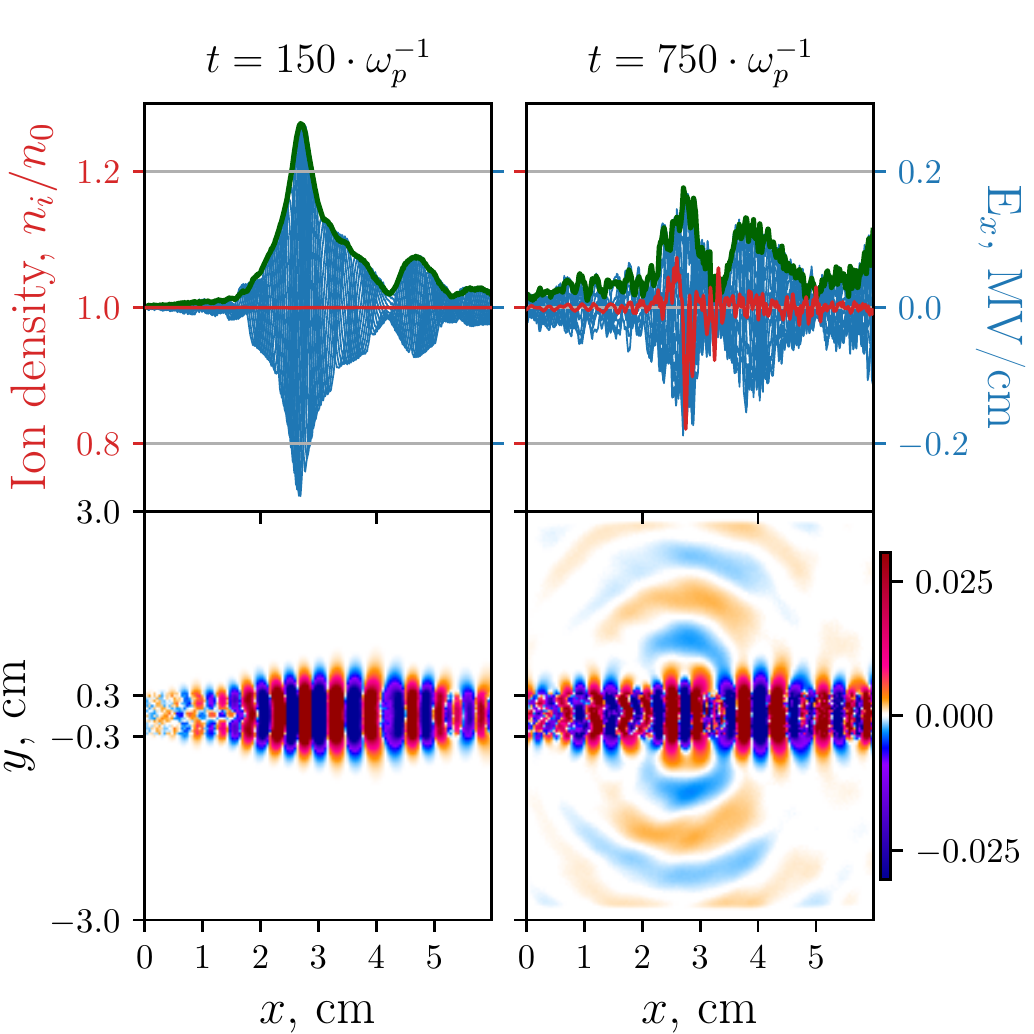} \caption{Top row: plasma wave amplitude $E_0(x)$ (green curves), longitudinal electric field $E_x(x)$ in the center of the plasma (different phases of one oscillation period are shown by blue curves) and ion density profile (red curves) averaged over the plasma thickness  and over 10 cells in the longitudinal direction in the case of $\Omega_e=1.22\omega_p$ in two moments of time. Bottom row: corresponding maps of longitudinal electric field $E_x(x,y)$ in MV/cm. }\label{pic:list1}
\end{figure}
\begin{figure*}[htb]
	\includegraphics[width=\linewidth]{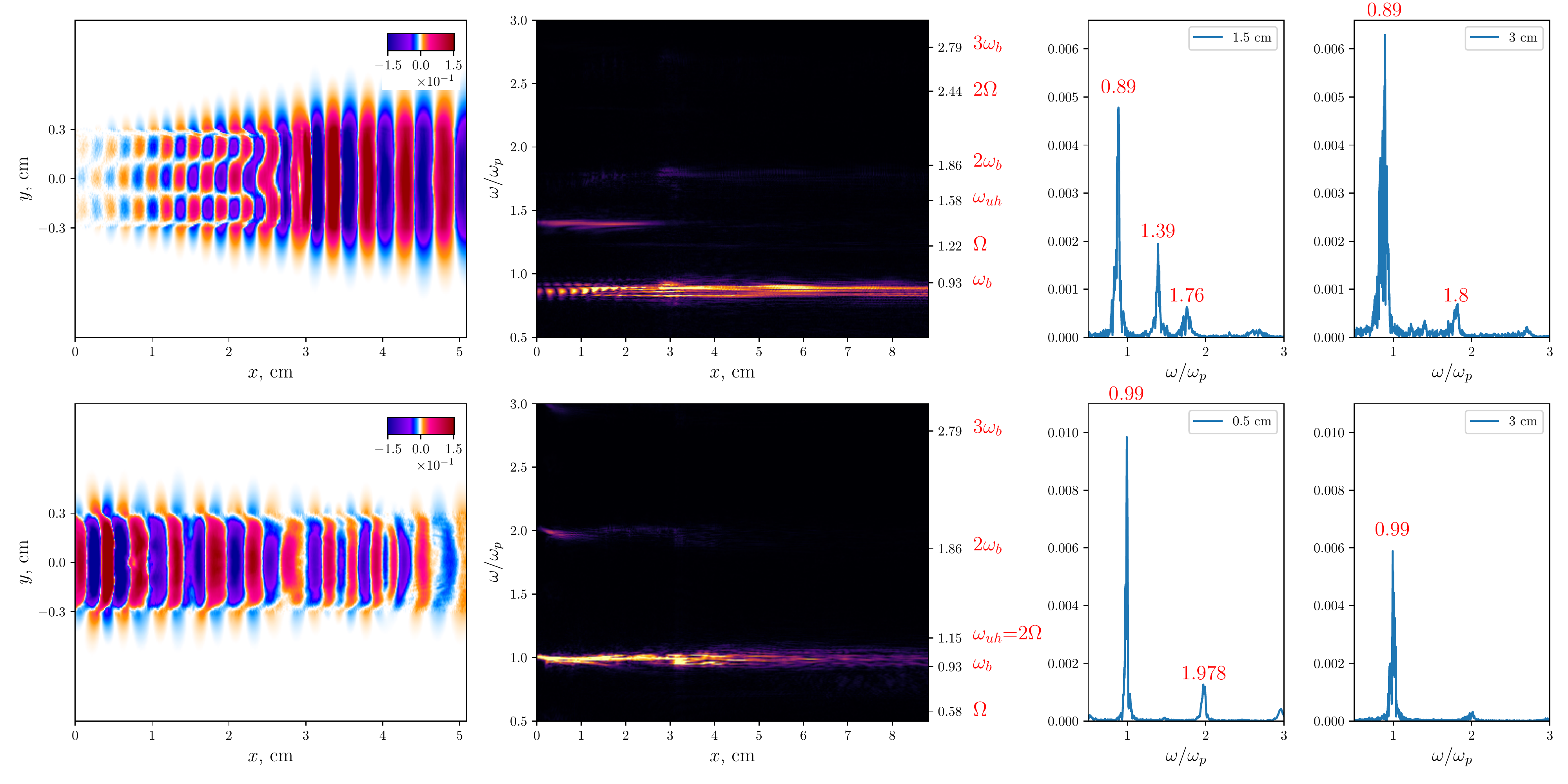} \caption{Top row: $\Omega_e=1.22\omega_p$, bottom row:  $\Omega_e=0.577\omega_p$. From left to right: the map of electric field $E_x$ inside the plasma in the  moment $t=300\,\omega_p^{-1}$, the Fourier spectrum of $E_x(t,x)$ in the center of a plasma column as a function of $x$, frequency spectra at individual spatial $x$-points. All spectra are calculated for the full duration of the simulation.}\label{pic:PlasmaFFT}
\end{figure*}

\subsection{Simulation results}

Let us carry out simulations of continuous electron beam injection into a magnetized plasma  for two values of the external magnetic field: $B=1.24$ T ($\Omega_e/\omega_p=1.22$) and  $B=0.586$ T ($\Omega_e/\omega_p=0.577$), where $\Omega_e=eB/m_e c$ is the electron cyclotron frequency. The former corresponds to the  conditions of GOL-3 experiments\cite{Burdakov2013}, while the latter -- to the case when the upper hybrid frequency equals to the doubled electron cyclotron frequency $\omega_{uh}=\sqrt{\omega_p^2+\Omega_e^2}=2\Omega_e$. Further, we will conventionally call these fields weak and strong ones.

First, consider the evolution of plasma oscillations at the early stages of the beam-plasma instability. For both values of the external magnetic field, the most intense beam-plasma interaction is localized at a distance of less than $6$ cm from the injector. The linear stage of the two-stream instability is followed by the nonlinear process of  beam trapping by the field of an excited wave. As a result, spatially localized wave packets with a large amplitude of the longitudinal electric field are formed (Fig. \ref{pic:list1} $t=150\,\omega_p^{-1}$).

In the strong magnetic field, the role of the most unstable modes are played by oblique oscillations of the upper-hybrid branch getting in the Cherenkov resonance with the beam and having the frequency $1.39\omega_p$ (Fig. \ref{pic:PlasmaFFT} top). These oscillations are excited in the first 3 cm of the plasma. After interacting with them, the beam receives a significant spread in momenta and begins to build up purely longitudinal waves with the frequency $0.89\omega_p$. It is slightly lower than the value of $\omega_b = 0.93\omega_p$ that is predicted by the linear theory (\ref{fr}) for a monoenergetic beam. As the plasma electrons are heated, the oblique instability decays and the beam begins to build up longitudinal oscillations in the entire region.

In the weak magnetic field $\Omega_e/\omega_p = 0.577$, the beam excites only longitudinal plasma oscillations at the frequency $0.99\,\omega_p$ from the very beginning. Since the angular spread of the beam in this case turns out to be noticeably smaller, the instability growth rate increases significantly, shifting the region of intense beam relaxation to the immediate vicinity of the injector (Fig. \ref{pic:PlasmaFFT} bottom).

\begin{figure*}[htb]
	\includegraphics[width=\linewidth]{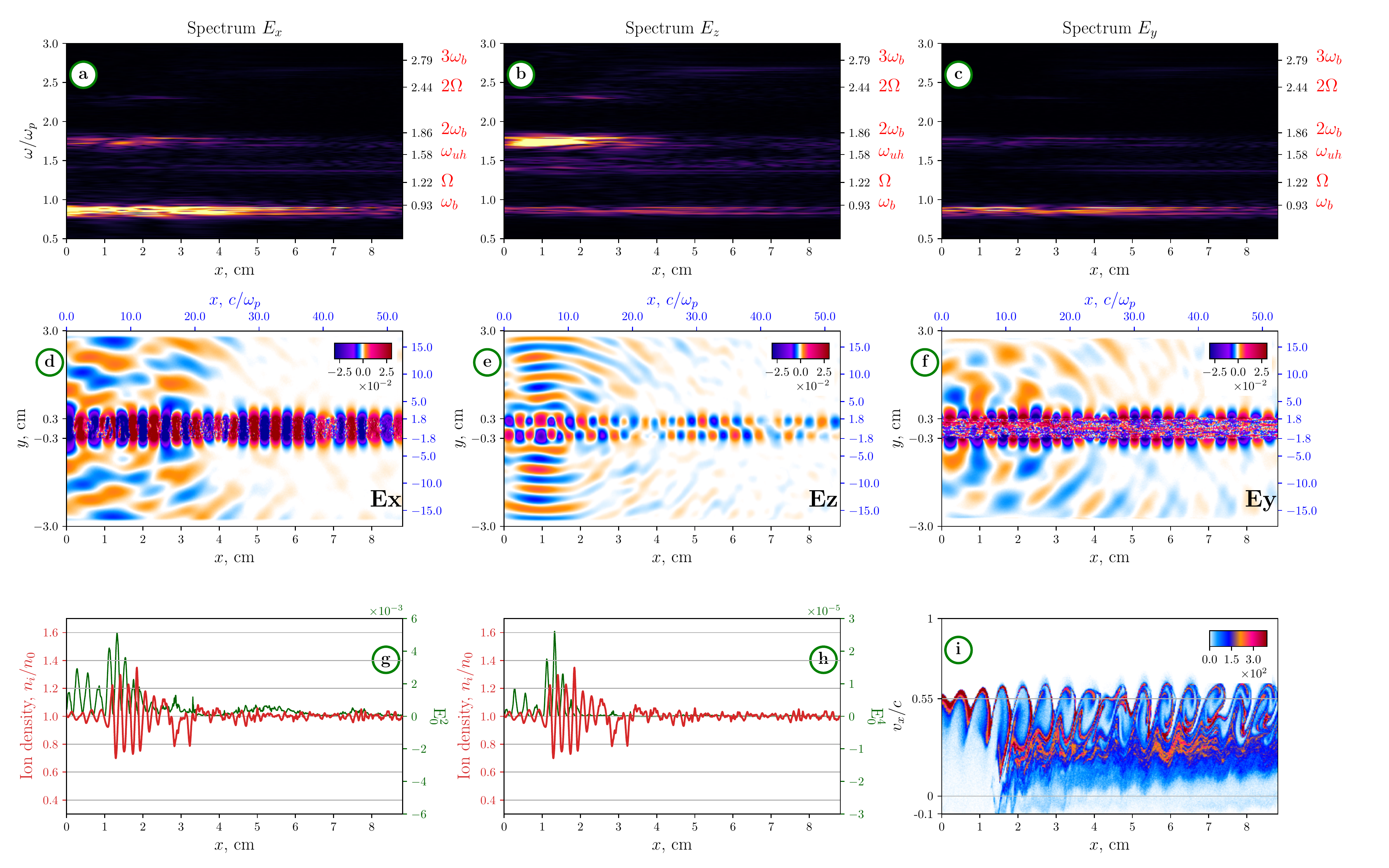} \caption{Simulation results for the magnetic field $\Omega=1.22\omega_p$. Upper row (a-c): radiation spectra of electric field components for the entire simulation time at the outer boundary of the vacuum region as  functions of the longitudinal coordinate $x$; middle row (d-f): maps of electric fields in the moment $t=1314\,\omega_p^{-1}=7.37$ ns; bottom row: (g-f) the plasma ion density averaged over the plasma thickness  and over 10 cells in the longitudinal direction (red line); the square (g) and the fourth power (f) of the plasma wave amplitude $E_0(x)$ in the center of the plasma (green line); (i) phase space  $(x, v_x)$ of the beam.}\label{pic:Om122all}
\end{figure*}

\begin{figure}[htb]
	\includegraphics[width=\linewidth]{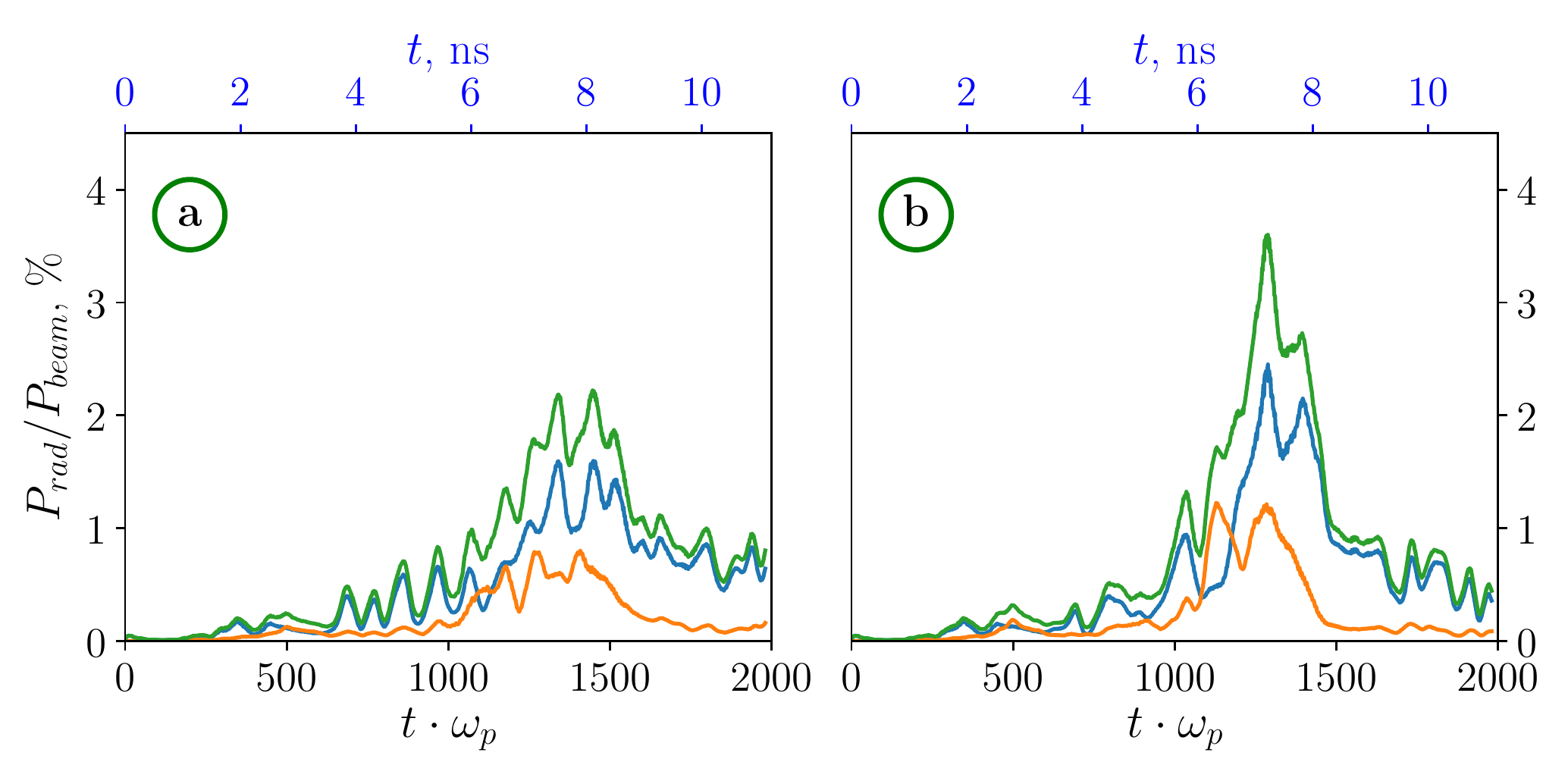} 
	\caption{The efficiency of beam-to-radiation power conversion as a function of time in the strong magnetic field $\Omega_e = 1.22\omega_p$ for two simulations with different random realizations of initial particle distributions. The blue line is TM $(E_x, E_y, B_z)$ mode, the orange line is TE $(B_x, B_y, E_z)$ mode, the green line is their sum.}
	\label{pic:eff122}
\end{figure}

In further, the ponderomotive force of the spatially localized wave packet forms a well on the longitudinal profile of the ion density. The linear mode conversion of plasma waves on this perturbation gives rise to first microbursts of EM radiation near the plasma frequency (Fig. \ref{pic:list1} $t=750\,\omega_p^{-1}$). At the same time, both in the density profile and in the amplitude of plasma oscillations, we observe a growth of the smaller-scale modulation instability with $q=2k_{\|}$. The nonlinear stage of this instability ends with the formation of a deep longitudinal modulation of the ion density and transformation of traveling plasma oscillations into a standing wave whose antinodes are localized inside the plasma density wells (Fig. \ref{pic:Om122all} g-h). While oscillations in neighboring wells occur in antiphase (i.e. the standing wave consists of traveling waves with $\pm k_{\|}$), they are pumped resonantly by the beam.  According to the plasma antenna mechanism, the nonlinear interaction of modes with $\pm k_{\|}$ leads to transverse emission at $2\omega_b$ (Fig. \ref{pic:Om122all} e), and their linear conversion on perturbations with $q = k_{\|}$ is responsible for radiation at $\omega_b$ (Fig. \ref {pic:Om122all} d). As neighboring oscillators dephase, they fall out of resonance with the beam and the locked plasma oscillations begin to decay. Thus, the modulational instability of the most unstable beam-driven wave with subsequent trapping of oscillations in density wells and the local disruption of the instability forms a single burst of EM radiation. Maps of the electric fields in the moment of the most intense generation of EM waves in the strong magnetic field are presented in Fig. \ref{pic:Om122all} d-f, while the temporal evolution of a single burst is shown in Fig. \ref{pic:eff122} a.

\begin{figure*}[htb]
	\includegraphics[width=\linewidth]{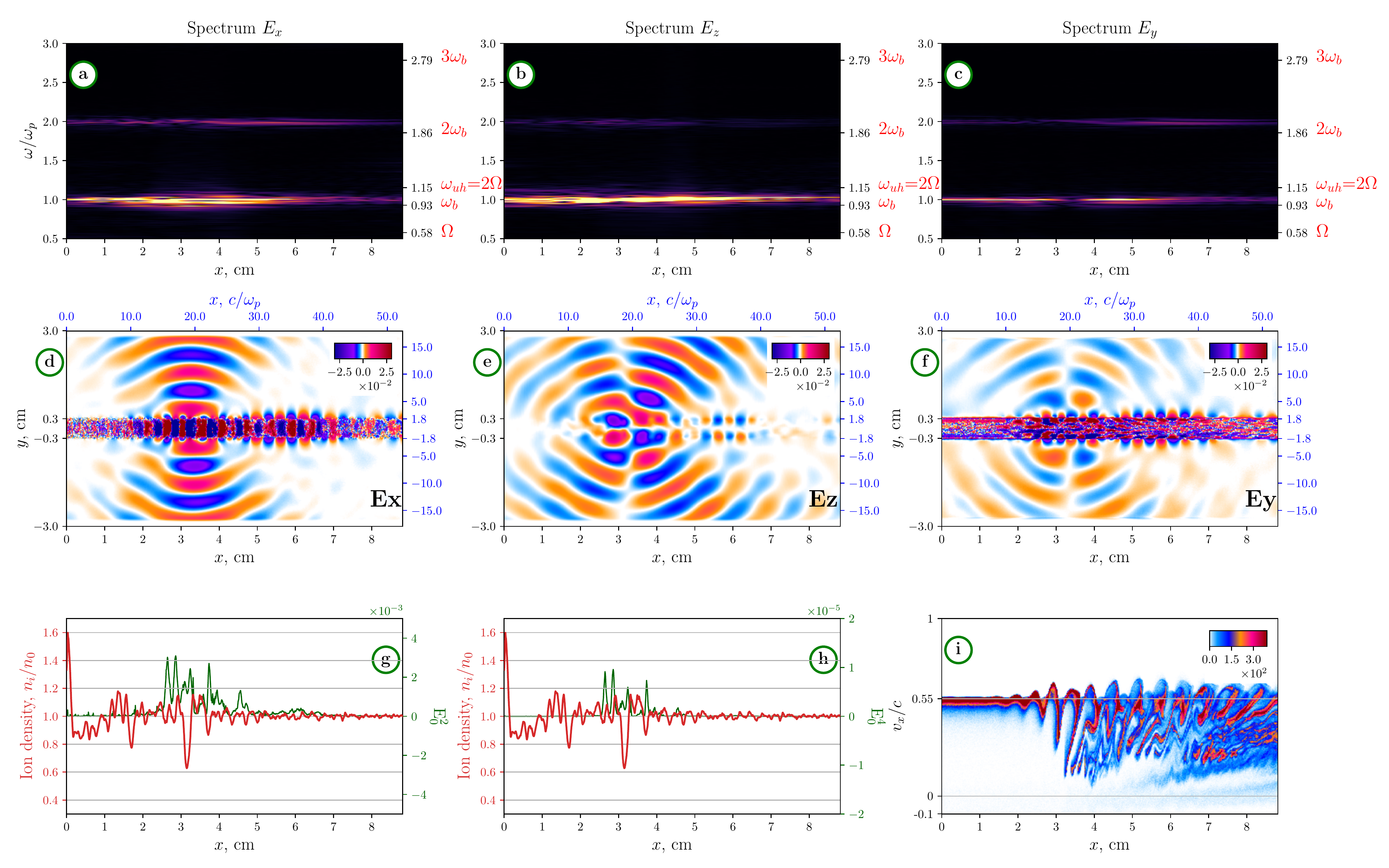}
	\caption{Same as in Fig. \ref{pic:Om122all}, but for the magnetic field $\Omega = 0.577\omega_p$ and time moment $t = 1910\,\omega_p^{-1}= 10.7$ ns.} \label{pic:Om057all}
	\end{figure*}

Figure \ref{pic:Om122all} (d,e,g,h) also shows that the fundamental EM emission is less localized than the second harmonic one, which is consistent with our theoretical scalings $P_{\omega_b}\propto E_0^2$  and $P_{2\omega_b}\propto E_0^4$. An interesting feature of the considered regime of beam-plasma interaction is that emissions at different harmonics (Fig. \ref{pic:Om122all} a-c) differ in polarizations: the spectrum of the TM $(E_x, E_y, B_z)$ mode is dominated by the beam frequency $\omega_b$, while in the TE $(B_x, B_y, E_z)$ mode, most of the radiation energy is concentrated at the second harmonic. The radiation spectrum also demonstrates the less intensive lines near $\omega = 1.39\omega_p$ and $\omega\approx2\Omega_e$ indicating that resonant upper hybrid oscillations and harmonic cyclotron emissions are able to escape from the plasma.

Note that the formation of a nonuniform ion density profile depends on a complicated history of the development of the two-stream and modulational instabilities and, for each random realization of the particle distribution function, takes place in its own unique way. In this situation, only the global scenario of the process, the spectrum of the observed radiation, as well as the conclusion about its high generation efficiency remain unchanged. Figure \ref{pic:eff122} presents the efficiency of the beam-to-radiation power conversion in the case of the strong magnetic field for two simulations differing only in the specific implementation of the initial momentum distribution function of plasma particles. It can be seen that the total radiation power constitutes  2\% - 3.5\% of the injected beam power, while the efficiency of the $2\omega_p$-radiation polarized across the magnetic field reaches 1\%.

\begin{figure}[htb]
	\includegraphics[width=\linewidth]{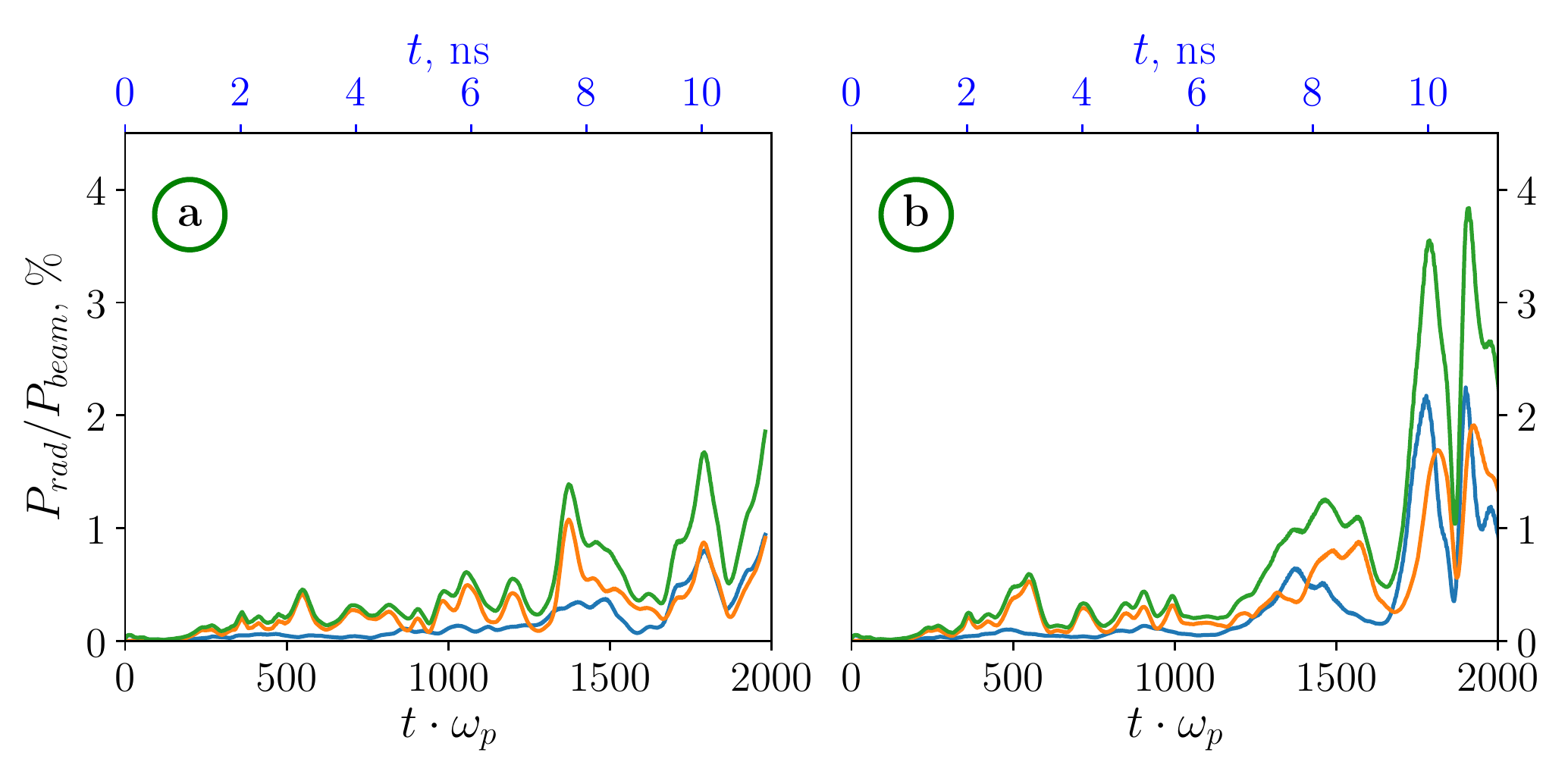} 
	\caption{Same as in Fig. \ref{pic:eff122}, but for the case $\Omega_e=0.577\omega_p$. }
		\label{pic:eff057}
\end{figure}

Results of simulations for the weak magnetic field $\Omega_e = 0.577\omega_p $ in the moment of the most efficient EM emission are shown in Fig. \ref{pic:Om057all}. The corresponding temporal dependence of the relative radiation power is still a set of bursts (Fig. \ref{pic:eff057} b).
Initially, the most intense beam-plasma interaction was localized in the immediate vicinity of the injector. This contributed to the formation of a small-scale ion density modulation there. As can be seen from Fig. \ref{pic:Om057all} d and g, this modulation led to a complete disruption of the instability in this region. As a result, the beam began to pump oscillations effectively in a more remote area (3-4 cm) which eventually also became a source of EM emission.

\begin{figure}[htb]
	\includegraphics[width=\linewidth]{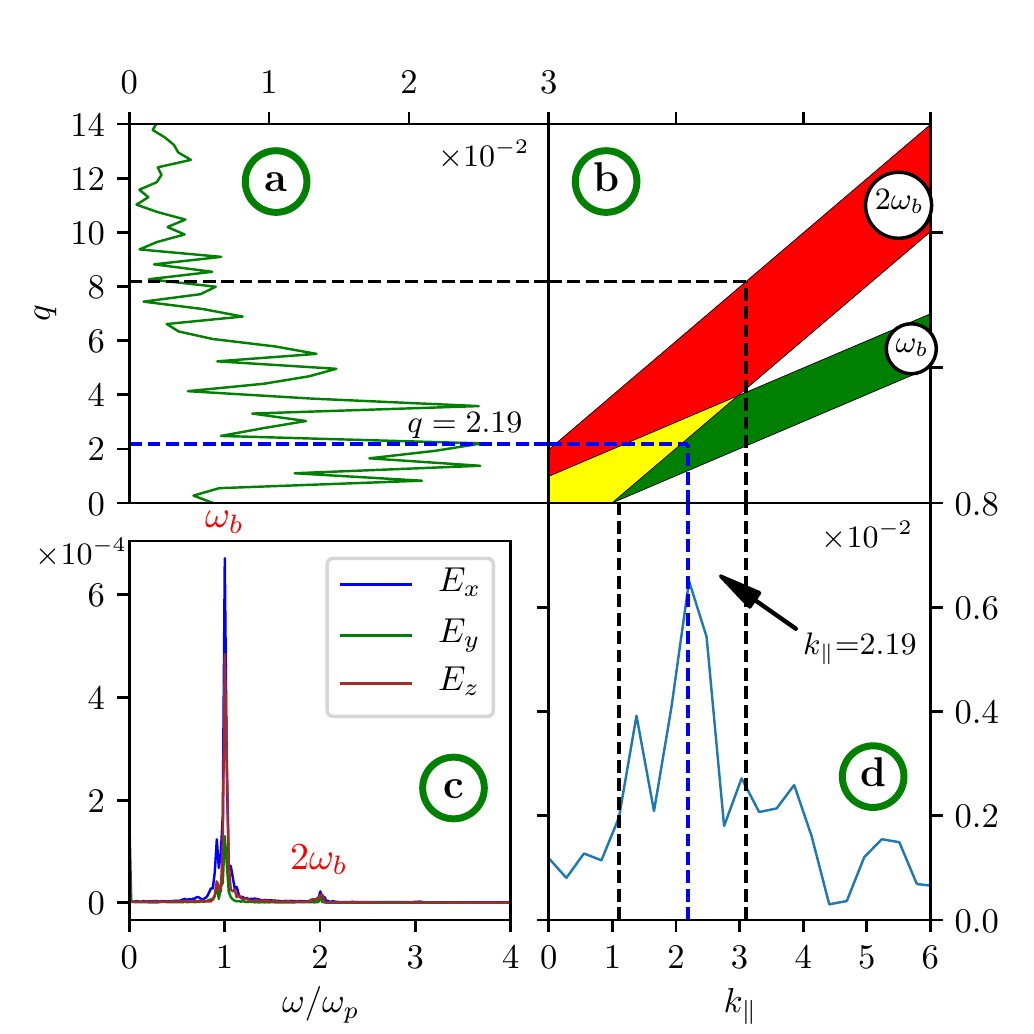} 
	\caption{Conditions of EM waves generation via the antenna mechanism for the case $\Omega_e=0.577\omega_p$. (a) Ion density spectrum in the moment $t=1910\,\omega_p^{-1}=10.7$ ns; (d) spectrum of the plasma wave in the center of the plasma layer; (b) conditions of efficient realization of the plasma antenna mechanism in ($q,\,k_\parallel$) coordinates; (c) emission spectra at a point in front of the emitting zone.}\label{pic:Om055ant}
\end{figure}

Let us analyze the role of the antenna mechanism in the generation of such radiation. For this purpose,  we consider the spectra of ion density (fig. \ref{pic:Om055ant} a) and plasma oscillations (fig. \ref{pic:Om055ant} d) in the same moment of time that is shown in Fig. \ref{pic:Om057all}. In addition, we present the emission spectrum (Fig. \ref {pic:Om055ant} c) in a single spatial point near the absorbing layer lying oppositely the radiating region and also indicate areas in the ($q,k_\parallel$)-space (Fig. \ref{pic:Om055ant} b) in which the conditions (\ref{Anwp}) and (\ref {An2wp}) for antenna emissions are satisfied ($v_b=0.99\omega_p/k_\parallel$). 

In the spectrum of plasma oscillations, it is observed the dominance of the beam mode with the wavenumber $k_\parallel = 2.19$ that is coincides with one of the main peaks of the ion density spectrum $q = 2.19$. This situation corresponds to the transverse antenna emission near the plasma frequency. For the remaining  spectral lines, it is also possible to generate EM waves at an angle to the plasma layer at both $\omega_p$ harmonics, which is confirmed by the electric field map and the emission spectrum shown in Fig. \ref{pic:Om057all}. In contrast to the case of the strong magnetic field $\Omega_e = 1.22\omega_p $,  the fundamental EM emission here dominates in all polarizations. Figure \ref {pic:eff057} shows the hystory of radiation efficiency for two different simulations with the magnetic field $\Omega_e = 0.577\omega_p$. For each polarization, the radiation power also reaches 1\%-2\% of the beam power.

\section{Disruption of the beam-plasma instability}\label{sec:Grad}

Presented simulations  show the possibility of generating EM waves by an electron beam in an initially uniform plasma channel with the high efficiency of power conversion (a few \%). But compact radiating  regions turn out to be located at a distance of several centimeters from the injector, which cannot explain experimental observations of efficient $2\omega_p$-emission at a distance of 84 cm. As it has been already discussed  for the case $\Omega=0.577\omega_p$, small-scale ion density perturbations  growing due to the modulational instability enable to completely  disrupt the two-stream instability. As a result, the beam becomes capable of pumping plasma waves in regions more distant from the injector. It is logically to assume that, in the future, a highly inhomogeneous density profile will be formed also in these new regions forcing  the beam to relax further and further away from the injector.  Such a scenario could explain the wider region of intense beam-plasma interaction  observed in laboratory experiments on microsecond time-scales.

\begin{figure}[htb]
	\includegraphics[width=\linewidth]{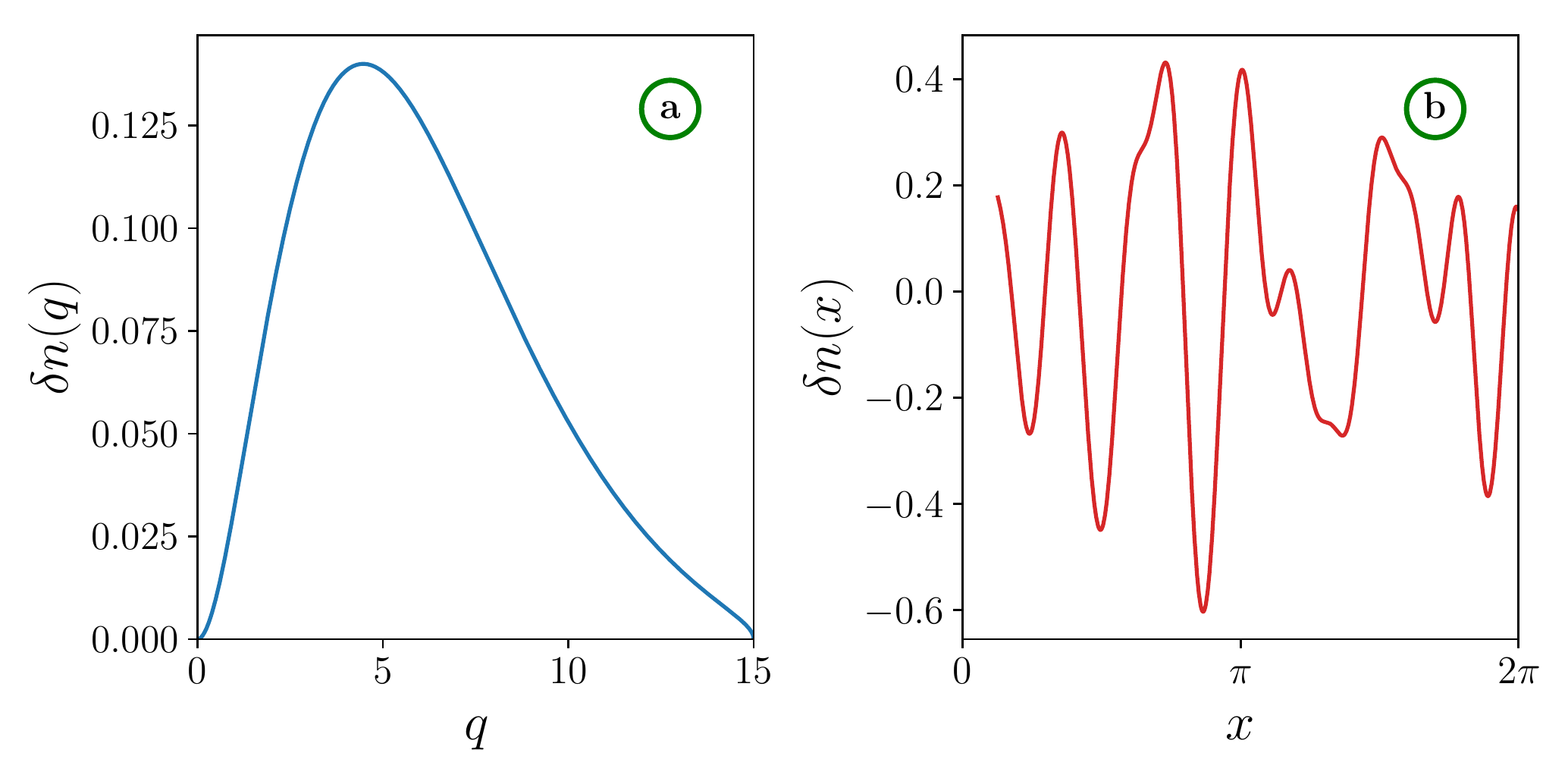} \caption{(a) The wavenumber spectrum of model density modulation $\delta n(q)$. (b) One period of a model density modulation $\delta n(x)$.}\label{pic:GammaDistr}
\end{figure}

To verify this hypothesis, we carry out simulations with a preformed turbulent ion density profile. The following function is taken as a model spectrum of density fluctuations (Fig. \ref{pic:GammaDistr} a):
\begin{equation}
f(q)=\dfrac{q^{k-1}}{\theta^k\Gamma(k)}e^{-q/\theta}\sqrt{q_{max}-q},
\end{equation}
where $q$ is the wavenumber of the longitudinal density modulation, $\theta=2.5$ and $k=3$ are distribution parameters, $q_{max}$ ia a maximal considered harmonic. 
From this distribution, $N_q=15$ uniformly arranged harmonics are selected.  For each of them,  we calculate the resulting amplitude and random phase in the range $[-\pi:\pi]$:
\begin{align}
\delta n_i&=\delta n_0\cdot \dfrac{f(q_i)}{f^{max}},\\
\phi_i&=\pi\cdot random[-1:1],
\end{align}
where $f^{max}$ is a maximum value of the distribution function and $\delta n_0=0.14$. The final plasma density is created in the form (Fig. \ref{pic:GammaDistr} b):
\begin{equation}
n(x)=n_0+\sum_{i=0}^{N_q}\delta n_i\sin(q_i x+\phi_i)
\end{equation}

\begin{figure*}[htb]
	\includegraphics[width=\linewidth]{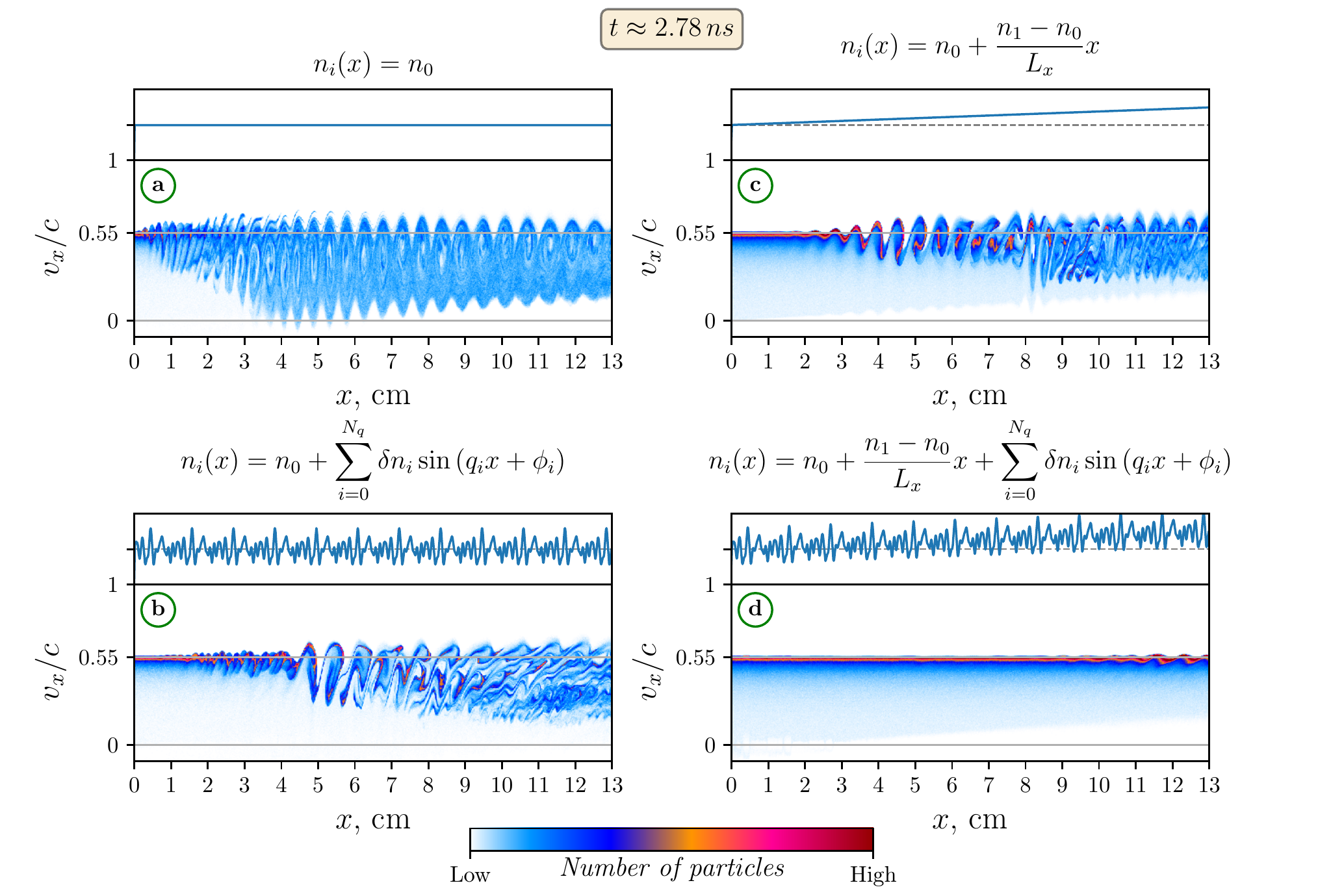} \caption{Each subfigure contains: the top -- ion density formula, in the middle -- density profile as a function of the longitudinal coordinate, bottom -- phase portrait of the beam ($x,v_x$). (a) -- homogeneous density; (b) -- turbulent plasma density superimposed on a uniform profile; (c) -- large-scale linear density gradient; (d) -- turbulent plasma density superimposed on a large-scale linear density gradient.}\label{pic:Sriv}
\end{figure*}

Figure \ref{pic:Sriv}a shows the phase plane of the beam $(x,v_x)$ in the case of its injection into a homogeneous plasma with fixed ions in the magnetic field $\Omega_e=1.22\omega_p$. The beam-plasma instability is seen  to be developed in the immediate vicinity of the injection area and, at a distance of 5 cm, beam-driven plasma oscillations reach amplitudes capable of stopping beam particles.

In the case of turbulent density (Fig. \ref{pic:Sriv} b), we observe that the beam-plasma instability is significantly weakened  in the first centimeters of plasma, but oscillations still reach the nonlinear level of beam trapping further from the injector. Thus, the presence of  small-scale density perturbations in a plasma with a sufficiently wide spectrum results in only weakening the instability, but not in its complete disruption.

Another possible explanation for the ability of the beam to generate EM radiation far from the injector is the presence of a large-scale longitudinal density gradient caused by the nonuniform gas inlet  into the vacuum chamber. To test this idea, we simulate the beam injection into a plasma with the density:
$$n(x)=n_0+\dfrac{n_1-n_0}{L_x} x,$$ where $n_0=10^{13}$ cm$^{-3}$, $n_1=10^{14}$ cm$^{-3}$ and $L_x=100$ cm.
The result is shown in Figure \ref{pic:Sriv} c. The presence of such a gradient also weakens the beam-plasma instability, but does not lead to its complete disruption.

The presence of intense oscillations in a plasma with a longitudinal density gradient will eventually lead to the development of modulation instability and the formation of a turbulent density spectrum. So it is to study the influence of a large-scale gradient with small-scale turbulence on the beam relaxation: $$n(x)=n_0+\dfrac{n_1-n_0}{L_x}+\sum_{i=0}^{N_q}\delta n_i\sin(q_i x+\phi_i).$$
The result of simulation with such a plasma density is shown in Fig. \ref{pic:Sriv} d. It is seen that such an inhomogeneity is able to disrupt the instability on a scale of 10 cm.

\section{Scenario of experiment}\label{Experiment}

According to the simulation results, relaxation of an electron beam with an energy of 100 keV and a current of 75 A in the conditions of the laboratory experiment at the GOL-3 facility \cite{Burdakov2013}  should proceed according to the following scenario. The beam ionizes the gas in its cross section and, due to the development of the two-stream instability, excites longitudinal oscillations in the resulting plasma channel. Since the gas density  at the first meter of the trap varies by a factor of 10, the growth rate of this instability turns out to be  lower than in a homogeneous plasma. Therefore, plasma oscillations excited by the beam reach their nonlinear saturation at a distance of 10 cm. Due to the development of the modulational instability of the dominant beam-driven wave with $k_{\|}\approx \omega_p/v_b$, small-scale periodic perturbations of ion density with the wavenumbers $q\sim k_{\|}$ begin to grow. Due to the small transverse size of the system, comparable to the radiation wavelength, the linear and nonlinear conversion of the beam wave on these perturbations via the antenna mechanism leads to the emission of EM waves at the plasma frequency and its second harmonic. The region of intense beam-plasma interaction has a size of 1-3 cm, and the radiation from it has the bursty character. The duration of bursts is determined by the growth time of the modulational instability and takes 1-5 ns in our simulations. The repetition period of such bursts should be determined by the longer time of the ion density relaxation that recreates favorable conditions for the new buildup cycle of the beam-plasma instability. The peak radiation power in a single burst can be several percent of the beam power, and in a strong magnetic field $\Omega_e>\omega_p$, a significant part of this power is contained in the transversely polarized $2\omega_p$-radiation. After a local disruption of the beam-plasma instability, the region of intense beam relaxation  shifts away from the injector. Based on the performed simulations, it can be assumed that, for the beam current under study, the formation of small-scale perturbations on a large-scale density gradient leads to a shift of intense beam-plasma interaction to the region of 10-20 cm.

It should be noted that the considered regime with a current of 75 A does not allow direct comparison with the experiment in which, at a given current, no EM radiation has been detected \cite{Burdakov2013} at the location of radiometric diagnostics (84 cm from the injector). Highly efficient EM emission in this place has been observed only in a narrow range of weaker currents of 20-30 A. Since simulations of weak currents on a scale of 1 m goes beyond the computing resources available to us, we restrict ourselves to a qualitative assessment of the adequacy of the scenario that is observed in the regime already studied.

First, our simulations allow to answer the question of why radiation has not been registered at the location of 84 cm with beam currents above 75 A. Intense radiation at the harmonics of the plasma frequency in this case should be generated in the region of 10–20 cm. The decrease of the beam current and its relative density reduces the growth rate of the beam-plasma instability and increases the stabilizing effect of density inhomogeneities shifting the region of efficient beam-plasma interaction further from the injector. It is likely that, for $I_b=20-30$ A, this region comes to the visibility zone of the radiation detection system.

Second, our simulation studies confirm the principle possibility to generate radiation at the doubled plasma frequency with the efficiency of 1\% that is unusually high for a turbulent beam-plasma system. The beam-driven plasma antenna is proposed for the role of the mechanism capable of providing such an efficient EM emission.

Third, the established scenario makes it possible to interpret the bursty nature of radiation in experiments as a cyclic process in which the buildup of the two-stream instability in a very compact region of 1-3 cm is accompanied by the growth of longitudinal modulation of the ion density followed by getting out of resonance of trapped oscillations  with the beam and subsequent relaxation of small-scale density inhomogeneities. Taking into account the longer period of such a cycle at low currents, this process may be responsible for the generation of individual bursts with the typical duration of 20 ns observed in experiments \cite{Postupaev2013}.

\section{Summary} 

We have carried out PIC simulations of continuous injection of a thin sub-relativistic electron beam into a magnetized plasma channel with the parameters of a laboratory experiment at the GOL-3 mirror trap and have shown the principal possibility of generating EM radiation near the plasma frequency and its second harmonic with a total power of several percent of beam power. We have found regimes when a significant fraction of this power falls on the second harmonic, which is consistent with the estimates of the efficiency of such radiation in the experiment. The conducted studies shed light on the physics of beam relaxation in these experiments and allow a qualitative interpretation of their results. In particular, it has become clear that, even in a strong magnetic field, the radiation is tied to the plasma frequency harmonics, but not to the upper hybrid resonance or cyclotron frequencies. We have also confirmed that the plasma antenna mechanism can work efficiently under the experimental conditions, have described a scenario for the formation of individual radiation bursts and explained why radiations produced in the high-current regimes ($>$75 A) cannot be detected by existing diagnostics.

Note that the same regime of a thin beam can be realized in a denser plasma with a beam of millimeter diameter and kiloampere currents, which opens the prospect of equally effective generation of high-power radiation in the THz frequency range.

This work was supported by RFBR (Grant No.18-02-00232).


\begin{thebibliography}{40}%
	\makeatletter
	\providecommand \@ifxundefined [1]{%
		\@ifx{#1\undefined}
	}%
	\providecommand \@ifnum [1]{%
		\ifnum #1\expandafter \@firstoftwo
		\else \expandafter \@secondoftwo
		\fi
	}%
	\providecommand \@ifx [1]{%
		\ifx #1\expandafter \@firstoftwo
		\else \expandafter \@secondoftwo
		\fi
	}%
	\providecommand \natexlab [1]{#1}%
	\providecommand \enquote  [1]{``#1''}%
	\providecommand \bibnamefont  [1]{#1}%
	\providecommand \bibfnamefont [1]{#1}%
	\providecommand \citenamefont [1]{#1}%
	\providecommand \href@noop [0]{\@secondoftwo}%
	\providecommand \href [0]{\begingroup \@sanitize@url \@href}%
	\providecommand \@href[1]{\@@startlink{#1}\@@href}%
	\providecommand \@@href[1]{\endgroup#1\@@endlink}%
	\providecommand \@sanitize@url [0]{\catcode `\\12\catcode `\$12\catcode
		`\&12\catcode `\#12\catcode `\^12\catcode `\_12\catcode `\%12\relax}%
	\providecommand \@@startlink[1]{}%
	\providecommand \@@endlink[0]{}%
	\providecommand \url  [0]{\begingroup\@sanitize@url \@url }%
	\providecommand \@url [1]{\endgroup\@href {#1}{\urlprefix }}%
	\providecommand \urlprefix  [0]{URL }%
	\providecommand \Eprint [0]{\href }%
	\providecommand \doibase [0]{http://dx.doi.org/}%
	\providecommand \selectlanguage [0]{\@gobble}%
	\providecommand \bibinfo  [0]{\@secondoftwo}%
	\providecommand \bibfield  [0]{\@secondoftwo}%
	\providecommand \translation [1]{[#1]}%
	\providecommand \BibitemOpen [0]{}%
	\providecommand \bibitemStop [0]{}%
	\providecommand \bibitemNoStop [0]{.\EOS\space}%
	\providecommand \EOS [0]{\spacefactor3000\relax}%
	\providecommand \BibitemShut  [1]{\csname bibitem#1\endcsname}%
	\let\auto@bib@innerbib\@empty
	\bibitem [{\citenamefont {Gurnett}\ and\ \citenamefont
		{Anderson}(1976)}]{Gurnett1976}%
	\BibitemOpen
	\bibfield  {author} {\bibinfo {author} {\bibfnamefont {D.~A.}\ \bibnamefont
			{Gurnett}}\ and\ \bibinfo {author} {\bibfnamefont {R.~R.}\ \bibnamefont
			{Anderson}},\ }\href {\doibase 10.1126/science.194.4270.1159} {\bibfield
		{journal} {\bibinfo  {journal} {Science}\ }\textbf {\bibinfo {volume}
			{194}},\ \bibinfo {pages} {1159} (\bibinfo {year} {1976})}\BibitemShut
	{NoStop}%
	\bibitem [{\citenamefont {Goldman}, \citenamefont {Reiter},\ and\ \citenamefont
		{Nicholson}(1980)}]{Goldman1980}%
	\BibitemOpen
	\bibfield  {author} {\bibinfo {author} {\bibfnamefont {M.~V.}\ \bibnamefont
			{Goldman}}, \bibinfo {author} {\bibfnamefont {G.~F.}\ \bibnamefont {Reiter}},
		\ and\ \bibinfo {author} {\bibfnamefont {D.~R.}\ \bibnamefont {Nicholson}},\
	}\href {\doibase 10.1063/1.862982} {\bibfield  {journal} {\bibinfo  {journal}
			{Physics of Fluids}\ }\textbf {\bibinfo {volume} {23}},\ \bibinfo {pages}
		{388} (\bibinfo {year} {1980})}\BibitemShut {NoStop}%
	\bibitem [{\citenamefont {Whelan}\ and\ \citenamefont
		{Stenzel}(1985)}]{Whelan1985}%
	\BibitemOpen
	\bibfield  {author} {\bibinfo {author} {\bibfnamefont {D.~A.}\ \bibnamefont
			{Whelan}}\ and\ \bibinfo {author} {\bibfnamefont {R.~L.}\ \bibnamefont
			{Stenzel}},\ }\href {\doibase 10.1063/1.865067} {\bibfield  {journal}
		{\bibinfo  {journal} {Physics of Fluids}\ }\textbf {\bibinfo {volume} {28}},\
		\bibinfo {pages} {958} (\bibinfo {year} {1985})}\BibitemShut {NoStop}%
	
	\bibitem{Pritchett1983}
	P.L. Pritchett and J.M. Dawson, Phys. Fluids {\bf 26}, 1114 (1983).
	
	\bibitem [{\citenamefont {Benford}\ \emph {et~al.}(1980)\citenamefont
		{Benford}, \citenamefont {Tzach}, \citenamefont {Kato},\ and\ \citenamefont
		{Smith}}]{Benford1980}%
	\BibitemOpen
	\bibfield  {author} {\bibinfo {author} {\bibfnamefont {G.}~\bibnamefont
			{Benford}}, \bibinfo {author} {\bibfnamefont {D.}~\bibnamefont {Tzach}},
		\bibinfo {author} {\bibfnamefont {K.}~\bibnamefont {Kato}}, \ and\ \bibinfo
		{author} {\bibfnamefont {D.~F.}\ \bibnamefont {Smith}},\ }\href {\doibase
		10.1103/PhysRevLett.45.1182} {\bibfield  {journal} {\bibinfo  {journal}
			{Physical Review Letters}\ }\textbf {\bibinfo {volume} {45}},\ \bibinfo
		{pages} {1182} (\bibinfo {year} {1980})}\BibitemShut {NoStop}%
	\bibitem [{\citenamefont {Whelan}\ and\ \citenamefont
		{Stenzel}(1981)}]{Whelan1981}%
	\BibitemOpen
	\bibfield  {author} {\bibinfo {author} {\bibfnamefont {D.~A.}\ \bibnamefont
			{Whelan}}\ and\ \bibinfo {author} {\bibfnamefont {R.~L.}\ \bibnamefont
			{Stenzel}},\ }\href {\doibase 10.1103/PhysRevLett.47.95} {\bibfield
		{journal} {\bibinfo  {journal} {Physical Review Letters}\ }\textbf {\bibinfo
			{volume} {47}},\ \bibinfo {pages} {95} (\bibinfo {year} {1981})}\BibitemShut
	{NoStop}%
	\bibitem [{\citenamefont {Cheung}\ \emph {et~al.}(1982)\citenamefont {Cheung},
		\citenamefont {Wong}, \citenamefont {Darrow},\ and\ \citenamefont
		{Qian}}]{Cheung1982}%
	\BibitemOpen
	\bibfield  {author} {\bibinfo {author} {\bibfnamefont {P.~Y.}\ \bibnamefont
			{Cheung}}, \bibinfo {author} {\bibfnamefont {A.~Y.}\ \bibnamefont {Wong}},
		\bibinfo {author} {\bibfnamefont {C.~B.}\ \bibnamefont {Darrow}}, \ and\
		\bibinfo {author} {\bibfnamefont {S.~J.}\ \bibnamefont {Qian}},\ }\href
	{\doibase 10.1103/PhysRevLett.48.1348} {\bibfield  {journal} {\bibinfo
			{journal} {Physical Review Letters}\ }\textbf {\bibinfo {volume} {48}},\
		\bibinfo {pages} {1348} (\bibinfo {year} {1982})}\BibitemShut {NoStop}%
	\bibitem [{\citenamefont {Che}\ \emph {et~al.}(2017)\citenamefont {Che},
		\citenamefont {Goldstein}, \citenamefont {Diamond},\ and\ \citenamefont
		{Sagdeev}}]{Che}%
	\BibitemOpen
	\bibfield  {author} {\bibinfo {author} {\bibfnamefont {H.}~\bibnamefont
			{Che}}, \bibinfo {author} {\bibfnamefont {M.~L.}\ \bibnamefont {Goldstein}},
		\bibinfo {author} {\bibfnamefont {P.~H.}\ \bibnamefont {Diamond}}, \ and\
		\bibinfo {author} {\bibfnamefont {R.~Z.}\ \bibnamefont {Sagdeev}},\ }\href
	{\doibase 10.1073/pnas.1614055114} {\bibfield  {journal} {\bibinfo  {journal}
			{Proceedings of the National Academy of Sciences of the United States of
				America}\ }\textbf {\bibinfo {volume} {114}},\ \bibinfo {pages} {1502}
		(\bibinfo {year} {2017})}\BibitemShut {NoStop}%
	\bibitem [{\citenamefont {Li}\ and\ \citenamefont {Cairns}(2013)}]{Li2013}%
	\BibitemOpen
	\bibfield  {author} {\bibinfo {author} {\bibfnamefont {B.}~\bibnamefont
			{Li}}\ and\ \bibinfo {author} {\bibfnamefont {I.~H.}\ \bibnamefont
			{Cairns}},\ }\href {\doibase 10.1002/jgra.50445} {\bibfield  {journal}
		{\bibinfo  {journal} {Journal of Geophysical Research: Space Physics}\
		}\textbf {\bibinfo {volume} {118}},\ \bibinfo {pages} {4748} (\bibinfo {year}
		{2013})}\BibitemShut {NoStop}%
	\bibitem [{\citenamefont {Robinson}\ and\ \citenamefont
		{Cairns}(1998)}]{Robinson1998}%
	\BibitemOpen
	\bibfield  {author} {\bibinfo {author} {\bibfnamefont {P.~A.}\ \bibnamefont
			{Robinson}}\ and\ \bibinfo {author} {\bibfnamefont {I.~H.}\ \bibnamefont
			{Cairns}},\ }\href {\doibase 10.1023/A:1005018918391} {\bibfield  {journal}
		{\bibinfo  {journal} {Solar Physics}\ }\textbf {\bibinfo {volume} {181}},\
		\bibinfo {pages} {363} (\bibinfo {year} {1998})}\BibitemShut {NoStop}%
	\bibitem [{\citenamefont {Thejappa}\ and\ \citenamefont
		{MacDowall}(1998)}]{Thejappa1998}%
	\BibitemOpen
	\bibfield  {author} {\bibinfo {author} {\bibfnamefont {G.}~\bibnamefont
			{Thejappa}}\ and\ \bibinfo {author} {\bibfnamefont {R.~J.}\ \bibnamefont
			{MacDowall}},\ }\href {\doibase 10.1086/305526} {\bibfield  {journal}
		{\bibinfo  {journal} {$\backslash$Apj}\ }\textbf {\bibinfo {volume} {498}},\
		\bibinfo {pages} {465} (\bibinfo {year} {1998})}\BibitemShut {NoStop}%
	\bibitem [{\citenamefont {Thurgood}\ and\ \citenamefont
		{Tsiklauri}(2015)}]{Thurgood2015}%
	\BibitemOpen
	\bibfield  {author} {\bibinfo {author} {\bibfnamefont {J.~O.}\ \bibnamefont
			{Thurgood}}\ and\ \bibinfo {author} {\bibfnamefont {D.}~\bibnamefont
			{Tsiklauri}},\ }\href {\doibase 10.1051/0004-6361/201527079} {\bibfield
		{journal} {\bibinfo  {journal} {Astronomy {\&} Astrophysics}\ }\textbf
		{\bibinfo {volume} {584}},\ \bibinfo {pages} {A83} (\bibinfo {year}
		{2015})}\BibitemShut {NoStop}%
	\bibitem [{\citenamefont {Chernov}(2010)}]{Chernov2010}%
	\BibitemOpen
	\bibfield  {author} {\bibinfo {author} {\bibfnamefont {G.~P.}\ \bibnamefont
			{Chernov}},\ }\href {\doibase 10.1088/1674-4527/10/9/002} {\bibfield
		{journal} {\bibinfo  {journal} {Research in Astronomy and Astrophysics}\
		}\textbf {\bibinfo {volume} {10}},\ \bibinfo {pages} {821} (\bibinfo {year}
		{2010})}\BibitemShut {NoStop}%
	\bibitem [{\citenamefont {Kuznetsov}\ and\ \citenamefont
		{Vlasov}(2013)}]{Kuznetsov2013}%
	\BibitemOpen
	\bibfield  {author} {\bibinfo {author} {\bibfnamefont {A.}~\bibnamefont
			{Kuznetsov}}\ and\ \bibinfo {author} {\bibfnamefont {V.}~\bibnamefont
			{Vlasov}},\ }\href {\doibase 10.1016/J.PSS.2012.09.005} {\bibfield  {journal}
		{\bibinfo  {journal} {Planetary and Space Science}\ }\textbf {\bibinfo
			{volume} {75}},\ \bibinfo {pages} {167} (\bibinfo {year} {2013})}\BibitemShut
	{NoStop}%
	\bibitem [{\citenamefont {Postupaev}\ \emph {et~al.}(2011)\citenamefont
		{Postupaev}, \citenamefont {Arzhannikov}, \citenamefont {Astrelin},
		\citenamefont {Batkin}, \citenamefont {Burdakov}, \citenamefont {Burmasov},
		\citenamefont {Ivanov}, \citenamefont {Ivantsivsky}, \citenamefont {Kuklin},
		\citenamefont {Kuznetsov}, \citenamefont {Makarov}, \citenamefont {Mekler},
		\citenamefont {Polosatkin}, \citenamefont {Popov}, \citenamefont
		{Rovenskikh}, \citenamefont {Shoshin}, \citenamefont {Sinitsky},
		\citenamefont {Sklyarov}, \citenamefont {Sorokina}, \citenamefont {Sudnikov},
		\citenamefont {Sulyaev},\ and\ \citenamefont {Vyacheslavov}}]{Postupaev2011}%
	\BibitemOpen
	\bibfield  {author} {\bibinfo {author} {\bibfnamefont {V.}~\bibnamefont
			{Postupaev}}, \bibinfo {author} {\bibfnamefont {A.~V.}\ \bibnamefont
			{Arzhannikov}}, \bibinfo {author} {\bibfnamefont {V.}~\bibnamefont
			{Astrelin}}, \bibinfo {author} {\bibfnamefont {V.}~\bibnamefont {Batkin}},
		\bibinfo {author} {\bibfnamefont {A.~V.}\ \bibnamefont {Burdakov}}, \bibinfo
		{author} {\bibfnamefont {V.~S.}\ \bibnamefont {Burmasov}}, \bibinfo {author}
		{\bibfnamefont {I.}~\bibnamefont {Ivanov}}, \bibinfo {author} {\bibfnamefont
			{M.}~\bibnamefont {Ivantsivsky}}, \bibinfo {author} {\bibfnamefont
			{K.}~\bibnamefont {Kuklin}}, \bibinfo {author} {\bibfnamefont
			{S.}~\bibnamefont {Kuznetsov}}, \bibinfo {author} {\bibfnamefont
			{M.}~\bibnamefont {Makarov}}, \bibinfo {author} {\bibfnamefont
			{K.}~\bibnamefont {Mekler}}, \bibinfo {author} {\bibfnamefont
			{S.}~\bibnamefont {Polosatkin}}, \bibinfo {author} {\bibfnamefont
			{S.}~\bibnamefont {Popov}}, \bibinfo {author} {\bibfnamefont
			{A.}~\bibnamefont {Rovenskikh}}, \bibinfo {author} {\bibfnamefont
			{A.}~\bibnamefont {Shoshin}}, \bibinfo {author} {\bibfnamefont
			{S.}~\bibnamefont {Sinitsky}}, \bibinfo {author} {\bibfnamefont
			{V.}~\bibnamefont {Sklyarov}}, \bibinfo {author} {\bibfnamefont
			{N.}~\bibnamefont {Sorokina}}, \bibinfo {author} {\bibfnamefont
			{A.}~\bibnamefont {Sudnikov}}, \bibinfo {author} {\bibfnamefont
			{Y.}~\bibnamefont {Sulyaev}}, \ and\ \bibinfo {author} {\bibfnamefont
			{L.}~\bibnamefont {Vyacheslavov}},\ }\href {\doibase 10.13182/FST11-A11594}
	{\bibfield  {journal} {\bibinfo  {journal} {Fusion Science and Technology}\
		}\textbf {\bibinfo {volume} {59}},\ \bibinfo {pages} {144} (\bibinfo {year}
		{2011})}\BibitemShut {NoStop}%
	\bibitem [{\citenamefont {Arzhannikov}\ and\ \citenamefont
		{Timofeev}(2012)}]{Timofeev2012c}%
	\BibitemOpen
	\bibfield  {author} {\bibinfo {author} {\bibfnamefont {A.~V.}\ \bibnamefont
			{Arzhannikov}}\ and\ \bibinfo {author} {\bibfnamefont {I.~V.}\ \bibnamefont
			{Timofeev}},\ }\href {\doibase 10.1088/0741-3335/54/10/105004} {\bibfield
		{journal} {\bibinfo  {journal} {Plasma Physics and Controlled Fusion}\
		}\textbf {\bibinfo {volume} {54}},\ \bibinfo {pages} {105004} (\bibinfo
		{year} {2012})}\BibitemShut {NoStop}%
	\bibitem [{\citenamefont {Arzhannikov}\ \emph {et~al.}(2014)\citenamefont
		{Arzhannikov}, \citenamefont {Burdakov}, \citenamefont {Burmasov},
		\citenamefont {Gavrilenko}, \citenamefont {Ivanov}, \citenamefont {Kasatov},
		\citenamefont {Kuznetsov}, \citenamefont {Mekler}, \citenamefont
		{Polosatkin}, \citenamefont {Postupaev}, \citenamefont {Rovenskikh},
		\citenamefont {Sinitsky}, \citenamefont {Sklyarov},\ and\ \citenamefont
		{Vyacheslavov}}]{Arzhannikov2014}%
	\BibitemOpen
	\bibfield  {author} {\bibinfo {author} {\bibfnamefont {A.~V.}\ \bibnamefont
			{Arzhannikov}}, \bibinfo {author} {\bibfnamefont {A.~V.}\ \bibnamefont
			{Burdakov}}, \bibinfo {author} {\bibfnamefont {V.~S.}\ \bibnamefont
			{Burmasov}}, \bibinfo {author} {\bibfnamefont {D.~E.}\ \bibnamefont
			{Gavrilenko}}, \bibinfo {author} {\bibfnamefont {I.~A.}\ \bibnamefont
			{Ivanov}}, \bibinfo {author} {\bibfnamefont {A.~A.}\ \bibnamefont {Kasatov}},
		\bibinfo {author} {\bibfnamefont {S.~A.}\ \bibnamefont {Kuznetsov}}, \bibinfo
		{author} {\bibfnamefont {K.~I.}\ \bibnamefont {Mekler}}, \bibinfo {author}
		{\bibfnamefont {S.~V.}\ \bibnamefont {Polosatkin}}, \bibinfo {author}
		{\bibfnamefont {V.~V.}\ \bibnamefont {Postupaev}}, \bibinfo {author}
		{\bibfnamefont {A.~F.}\ \bibnamefont {Rovenskikh}}, \bibinfo {author}
		{\bibfnamefont {S.~L.}\ \bibnamefont {Sinitsky}}, \bibinfo {author}
		{\bibfnamefont {V.~F.}\ \bibnamefont {Sklyarov}}, \ and\ \bibinfo {author}
		{\bibfnamefont {L.~N.}\ \bibnamefont {Vyacheslavov}},\ }\href {\doibase
		10.1063/1.4891884} {\bibfield  {journal} {\bibinfo  {journal} {Physics of
				Plasmas}\ }\textbf {\bibinfo {volume} {21}},\ \bibinfo {pages} {082106}
		(\bibinfo {year} {2014})}\BibitemShut {NoStop}%
	\bibitem{Arzhannikov2016}
	A.V. Arzhannikov, A.V. Burdakov, V.S. Burmasov, I.A. Ivanov, A.A.
Kasatov, S.A. Kuznetsov, M.A. Makarov, K.I. Mekler, S.V. Polosatkin,
S.S. Popov, V.V. Postupaev, A.F. Rovenskikh, S.L. Sinitsky, V.F.
Sklyarov, V.D. Stepanov, I.V. Timofeev, and M.K.A. Thumm, IEEE
Trans. Terahertz Sci. Technol. {\bf 6}, 245 (2016).
  \bibitem [{\citenamefont {Burdakov}\ \emph {et~al.}(2013)\citenamefont
		{Burdakov}, \citenamefont {Arzhannikov}, \citenamefont {Burmasov},
		\citenamefont {Ivanov}, \citenamefont {Ivantsivsky}, \citenamefont
		{Kandaurov}, \citenamefont {Kuznetsov}, \citenamefont {Kurkuchekov},
		\citenamefont {Mekler}, \citenamefont {Polosatkin}, \citenamefont {Popov},
		\citenamefont {Postupaev}, \citenamefont {Rovenskikh}, \citenamefont
		{Sklyarov}, \citenamefont {Thumm}, \citenamefont {Trunev},\ and\
		\citenamefont {Vyacheslavov}}]{Burdakov2013}%
	\BibitemOpen
	\bibfield  {author} {\bibinfo {author} {\bibfnamefont {A.~V.}\ \bibnamefont
			{Burdakov}}, \bibinfo {author} {\bibfnamefont {A.~V.}\ \bibnamefont
			{Arzhannikov}}, \bibinfo {author} {\bibfnamefont {V.~S.}\ \bibnamefont
			{Burmasov}}, \bibinfo {author} {\bibfnamefont {I.~A.}\ \bibnamefont
			{Ivanov}}, \bibinfo {author} {\bibfnamefont {M.~V.}\ \bibnamefont
			{Ivantsivsky}}, \bibinfo {author} {\bibfnamefont {I.~V.}\ \bibnamefont
			{Kandaurov}}, \bibinfo {author} {\bibfnamefont {S.~A.}\ \bibnamefont
			{Kuznetsov}}, \bibinfo {author} {\bibfnamefont {V.~V.}\ \bibnamefont
			{Kurkuchekov}}, \bibinfo {author} {\bibfnamefont {K.~I.}\ \bibnamefont
			{Mekler}}, \bibinfo {author} {\bibfnamefont {S.~V.}\ \bibnamefont
			{Polosatkin}}, \bibinfo {author} {\bibfnamefont {S.~S.}\ \bibnamefont
			{Popov}}, \bibinfo {author} {\bibfnamefont {V.~V.}\ \bibnamefont
			{Postupaev}}, \bibinfo {author} {\bibfnamefont {A.~F.}\ \bibnamefont
			{Rovenskikh}}, \bibinfo {author} {\bibfnamefont {V.~F.}\ \bibnamefont
			{Sklyarov}}, \bibinfo {author} {\bibfnamefont {M.~K.~A.}\ \bibnamefont
			{Thumm}}, \bibinfo {author} {\bibfnamefont {Y.~A.}\ \bibnamefont {Trunev}}, \
		and\ \bibinfo {author} {\bibfnamefont {L.~N.}\ \bibnamefont {Vyacheslavov}},\
	}\href {\doibase 10.13182/FST13-A16930} {\bibfield  {journal} {\bibinfo
			{journal} {Fusion Science and Technology}\ }\textbf {\bibinfo {volume}
			{63}},\ \bibinfo {pages} {286} (\bibinfo {year} {2013})}\BibitemShut
	{NoStop}%
	\bibitem{Postupaev2013}
	V.V. Postupaev, A.V. Burdakov, I.A. Ivanov, V.F. Sklyarov, A.V.
Arzhannikov, D.Y. Gavrilenko, I.V. Kandaurov, A.A. Kasatov, V.V.
Kurkuchekov, K.I. Mekler, S.V. Polosatkin, S.S. Popov, A.F.
Rovenskikh, A.V. Sudnikov, Y.S. Sulyaev, Y.A. Trunev, and L.N.
Vyacheslavov, Phys. Plasmas {\bf 20}, 092304 (2013).
\bibitem [{\citenamefont {Ivanov}\ \emph {et~al.}(2015)\citenamefont {Ivanov},
		\citenamefont {Arzhannikov}, \citenamefont {Burdakov}, \citenamefont
		{Burmasov}, \citenamefont {Gavrilenko}, \citenamefont {Kasatov},
		\citenamefont {Kandaurov}, \citenamefont {Kurkuchekov}, \citenamefont
		{Kuznetsov}, \citenamefont {Mekler}, \citenamefont {Polosatkin},
		\citenamefont {Popov}, \citenamefont {Postupaev}, \citenamefont {Rovenskikh},
		\citenamefont {Sklyarov}, \citenamefont {Sorokina}, \citenamefont {Trunev},\
		and\ \citenamefont {Vyacheslavov}}]{Ivanov2015}%
	\BibitemOpen
	\bibfield  {author} {\bibinfo {author} {\bibfnamefont {I.~A.}\ \bibnamefont
			{Ivanov}}, \bibinfo {author} {\bibfnamefont {A.~V.}\ \bibnamefont
			{Arzhannikov}}, \bibinfo {author} {\bibfnamefont {A.~V.}\ \bibnamefont
			{Burdakov}}, \bibinfo {author} {\bibfnamefont {V.~S.}\ \bibnamefont
			{Burmasov}}, \bibinfo {author} {\bibfnamefont {D.~E.}\ \bibnamefont
			{Gavrilenko}}, \bibinfo {author} {\bibfnamefont {A.~A.}\ \bibnamefont
			{Kasatov}}, \bibinfo {author} {\bibfnamefont {I.~V.}\ \bibnamefont
			{Kandaurov}}, \bibinfo {author} {\bibfnamefont {V.~V.}\ \bibnamefont
			{Kurkuchekov}}, \bibinfo {author} {\bibfnamefont {S.~A.}\ \bibnamefont
			{Kuznetsov}}, \bibinfo {author} {\bibfnamefont {K.~I.}\ \bibnamefont
			{Mekler}}, \bibinfo {author} {\bibfnamefont {S.~V.}\ \bibnamefont
			{Polosatkin}}, \bibinfo {author} {\bibfnamefont {S.~S.}\ \bibnamefont
			{Popov}}, \bibinfo {author} {\bibfnamefont {V.~V.}\ \bibnamefont
			{Postupaev}}, \bibinfo {author} {\bibfnamefont {A.~F.}\ \bibnamefont
			{Rovenskikh}}, \bibinfo {author} {\bibfnamefont {V.~F.}\ \bibnamefont
			{Sklyarov}}, \bibinfo {author} {\bibfnamefont {N.~V.}\ \bibnamefont
			{Sorokina}}, \bibinfo {author} {\bibfnamefont {Y.~A.}\ \bibnamefont
			{Trunev}}, \ and\ \bibinfo {author} {\bibfnamefont {L.~N.}\ \bibnamefont
			{Vyacheslavov}},\ }\href {\doibase 10.1063/1.4936874} {\bibfield  {journal}
		{\bibinfo  {journal} {Physics of Plasmas}\ }\textbf {\bibinfo {volume}
			{22}},\ \bibinfo {pages} {122302} (\bibinfo {year} {2015})}\BibitemShut
	{NoStop}%
	\bibitem [{\citenamefont {Timofeev}, \citenamefont {Annenkov},\ and\
		\citenamefont {Arzhannikov}(2015)}]{Timofeev2015}%
	\BibitemOpen
	\bibfield  {author} {\bibinfo {author} {\bibfnamefont {I.~V.}\ \bibnamefont
			{Timofeev}}, \bibinfo {author} {\bibfnamefont {V.~V.}\ \bibnamefont
			{Annenkov}}, \ and\ \bibinfo {author} {\bibfnamefont {A.~V.}\ \bibnamefont
			{Arzhannikov}},\ }\href {\doibase 10.1063/1.4935890} {\bibfield  {journal}
		{\bibinfo  {journal} {Physics of Plasmas}\ }\textbf {\bibinfo {volume}
			{22}},\ \bibinfo {pages} {113109} (\bibinfo {year} {2015})}\BibitemShut
	{NoStop}%
	\bibitem [{\citenamefont {Annenkov}, \citenamefont {Volchok},\ and\
		\citenamefont {Timofeev}(2016)}]{Annenkov2016a}%
	\BibitemOpen
	\bibfield  {author} {\bibinfo {author} {\bibfnamefont {V.~V.}\ \bibnamefont
			{Annenkov}}, \bibinfo {author} {\bibfnamefont {E.~P.}\ \bibnamefont
			{Volchok}}, \ and\ \bibinfo {author} {\bibfnamefont {I.~V.}\ \bibnamefont
			{Timofeev}},\ }\href {\doibase 10.1088/0741-3335/58/4/045009} {\bibfield
		{journal} {\bibinfo  {journal} {Plasma Physics and Controlled Fusion}\
		}\textbf {\bibinfo {volume} {58}},\ \bibinfo {pages} {045009} (\bibinfo
		{year} {2016})}\BibitemShut {NoStop}%
	\bibitem [{\citenamefont {Timofeev}, \citenamefont {Volchok},\ and\
		\citenamefont {Annenkov}(2016)}]{Timofeev2016a}%
	\BibitemOpen
	\bibfield  {author} {\bibinfo {author} {\bibfnamefont {I.~V.}\ \bibnamefont
			{Timofeev}}, \bibinfo {author} {\bibfnamefont {E.~P.}\ \bibnamefont
			{Volchok}}, \ and\ \bibinfo {author} {\bibfnamefont {V.~V.}\ \bibnamefont
			{Annenkov}},\ }\href {\doibase 10.1063/1.4961218} {\bibfield  {journal}
		{\bibinfo  {journal} {Physics of Plasmas}\ }\textbf {\bibinfo {volume}
			{23}},\ \bibinfo {pages} {083119} (\bibinfo {year} {2016})}\BibitemShut
	{NoStop}%
	\bibitem [{\citenamefont {Annenkov}, \citenamefont {Timofeev},\ and\
		\citenamefont {Volchok}(2016{\natexlab{a}})}]{Annenkov2016b}%
	\BibitemOpen
	\bibfield  {author} {\bibinfo {author} {\bibfnamefont {V.~V.}\ \bibnamefont
			{Annenkov}}, \bibinfo {author} {\bibfnamefont {I.~V.}\ \bibnamefont
			{Timofeev}}, \ and\ \bibinfo {author} {\bibfnamefont {E.~P.}\ \bibnamefont
			{Volchok}},\ }\href {\doibase 10.1063/1.4948425} {\bibfield  {journal}
		{\bibinfo  {journal} {Physics of Plasmas}\ }\textbf {\bibinfo {volume}
			{23}},\ \bibinfo {pages} {053101} (\bibinfo {year}
		{2016}{\natexlab{a}})}\BibitemShut {NoStop}%
	\bibitem [{\citenamefont {Annenkov}\ \emph
		{et~al.}(2018{\natexlab{a}})\citenamefont {Annenkov}, \citenamefont
		{Berendeev}, \citenamefont {Volchok},\ and\ \citenamefont
		{Timofeev}}]{Annenkov2018a}%
	\BibitemOpen
	\bibfield  {author} {\bibinfo {author} {\bibfnamefont {V.~V.}\ \bibnamefont
			{Annenkov}}, \bibinfo {author} {\bibfnamefont {E.~A.}\ \bibnamefont
			{Berendeev}}, \bibinfo {author} {\bibfnamefont {E.~P.}\ \bibnamefont
			{Volchok}}, \ and\ \bibinfo {author} {\bibfnamefont {I.~V.}\ \bibnamefont
			{Timofeev}},\ }\href {http://arxiv.org/abs/1811.06426} {\  (\bibinfo {year}
		{2018}{\natexlab{a}})},\ \Eprint {http://arxiv.org/abs/1811.06426}
	{arXiv:1811.06426} \BibitemShut {NoStop}%
	\bibitem [{\citenamefont {Annenkov}, \citenamefont {Timofeev},\ and\
		\citenamefont {Volchok}(2016{\natexlab{b}})}]{Annenkov2016c}%
	\BibitemOpen
	\bibfield  {author} {\bibinfo {author} {\bibfnamefont {V.~V.}\ \bibnamefont
			{Annenkov}}, \bibinfo {author} {\bibfnamefont {I.~V.}\ \bibnamefont
			{Timofeev}}, \ and\ \bibinfo {author} {\bibfnamefont {E.~P.}\ \bibnamefont
			{Volchok}},\ }in\ \href {\doibase 10.1063/1.4964235} {\emph {\bibinfo
			{booktitle} {AIP Conference Proceedings}}},\ Vol.\ \bibinfo {volume} {1771}\
	(\bibinfo  {publisher} {AIP Publishing LLCAIP Publishing},\ \bibinfo {year}
	{2016})\ p.\ \bibinfo {pages} {070011}\BibitemShut {NoStop}%
	\bibitem [{\citenamefont {Ryutov}(1969)}]{Ryutov1969}%
	\BibitemOpen
	\bibfield  {author} {\bibinfo {author} {\bibfnamefont {D.}~\bibnamefont
			{Ryutov}},\ }\href {http://www.jetp.ac.ru/cgi-bin/dn/e_030_01_0131.pdf}
	{\bibfield  {journal} {\bibinfo  {journal} {Soviet Journal of Experimental
				and Theoretical Physics}\ }\textbf {\bibinfo {volume} {30}},\ \bibinfo
		{pages} {131} (\bibinfo {year} {1969})}\BibitemShut {NoStop}%
	\bibitem [{\citenamefont {Nishikawa}\ and\ \citenamefont
		{Ryutov}(1976)}]{Nishikawa1976}%
	\BibitemOpen
	\bibfield  {author} {\bibinfo {author} {\bibfnamefont {K.}~\bibnamefont
			{Nishikawa}}\ and\ \bibinfo {author} {\bibfnamefont {D.~D.}\ \bibnamefont
			{Ryutov}},\ }\href {\doibase 10.1143/JPSJ.41.1757} {\bibfield  {journal}
		{\bibinfo  {journal} {Journal of the Physical Society of Japan}\ }\textbf
		{\bibinfo {volume} {41}},\ \bibinfo {pages} {1757} (\bibinfo {year}
		{1976})}\BibitemShut {NoStop}%
	\bibitem [{\citenamefont {Krafft}\ \emph {et~al.}(2014)\citenamefont {Krafft},
		\citenamefont {Volokitin}, \citenamefont {Krasnoselskikh},\ and\
		\citenamefont {de~Wit}}]{Krafft2014}%
	\BibitemOpen
	\bibfield  {author} {\bibinfo {author} {\bibfnamefont {C.}~\bibnamefont
			{Krafft}}, \bibinfo {author} {\bibfnamefont {A.~S.}\ \bibnamefont
			{Volokitin}}, \bibinfo {author} {\bibfnamefont {V.~V.}\ \bibnamefont
			{Krasnoselskikh}}, \ and\ \bibinfo {author} {\bibfnamefont {T.~D.}\
			\bibnamefont {de~Wit}},\ }\href {\doibase 10.1002/2014JA020329} {\bibfield
		{journal} {\bibinfo  {journal} {Journal of Geophysical Research: Space
				Physics}\ }\textbf {\bibinfo {volume} {119}},\ \bibinfo {pages} {9369}
		(\bibinfo {year} {2014})}\BibitemShut {NoStop}%
	\bibitem [{\citenamefont {Pechhacker}\ and\ \citenamefont
		{Tsiklauri}(2014)}]{Pechhacker2014}%
	\BibitemOpen
	\bibfield  {author} {\bibinfo {author} {\bibfnamefont {R.}~\bibnamefont
			{Pechhacker}}\ and\ \bibinfo {author} {\bibfnamefont {D.}~\bibnamefont
			{Tsiklauri}},\ }\href {\doibase 10.1063/1.4863494} {\bibfield  {journal}
		{\bibinfo  {journal} {Physics of Plasmas}\ }\textbf {\bibinfo {volume}
			{21}},\ \bibinfo {pages} {012903} (\bibinfo {year} {2014})},\ \Eprint
	{http://arxiv.org/abs/1401.6966} {arXiv:1401.6966} \BibitemShut {NoStop}%
	\bibitem [{\citenamefont {Thurgood}\ and\ \citenamefont
		{Tsiklauri}(2016)}]{Thurgood2016}%
	\BibitemOpen
	\bibfield  {author} {\bibinfo {author} {\bibfnamefont {J.~O.}\ \bibnamefont
			{Thurgood}}\ and\ \bibinfo {author} {\bibfnamefont {D.}~\bibnamefont
			{Tsiklauri}},\ }\href {\doibase 10.1017/S0022377816000970} {\bibfield
		{journal} {\bibinfo  {journal} {Journal of Plasma Physics}\ }\textbf
		{\bibinfo {volume} {82}},\ \bibinfo {pages} {905820604} (\bibinfo {year}
		{2016})},\ \Eprint {http://arxiv.org/abs/1612.01780} {arXiv:1612.01780}
	\BibitemShut {NoStop}%
	\bibitem [{\citenamefont {Schmitz}\ and\ \citenamefont
		{Tsiklauri}(2013)}]{Schmitz2013}%
	\BibitemOpen
	\bibfield  {author} {\bibinfo {author} {\bibfnamefont {H.}~\bibnamefont
			{Schmitz}}\ and\ \bibinfo {author} {\bibfnamefont {D.}~\bibnamefont
			{Tsiklauri}},\ }\href {\doibase 10.1063/1.4812453} {\bibfield  {journal}
		{\bibinfo  {journal} {Physics of Plasmas}\ }\textbf {\bibinfo {volume}
			{20}},\ \bibinfo {pages} {062903} (\bibinfo {year} {2013})}\BibitemShut
	{NoStop}%
	\bibitem [{\citenamefont {Pechhacker}\ and\ \citenamefont
		{Tsiklauri}(2012)}]{Pechhacker2012}%
	\BibitemOpen
	\bibfield  {author} {\bibinfo {author} {\bibfnamefont {R.}~\bibnamefont
			{Pechhacker}}\ and\ \bibinfo {author} {\bibfnamefont {D.}~\bibnamefont
			{Tsiklauri}},\ }\href {\doibase 10.1063/1.4768429} {\bibfield  {journal}
		{\bibinfo  {journal} {Physics of Plasmas}\ }\textbf {\bibinfo {volume}
			{19}},\ \bibinfo {pages} {112903} (\bibinfo {year} {2012})},\ \Eprint
	{http://arxiv.org/abs/1211.6726} {arXiv:1211.6726} \BibitemShut {NoStop}%
	\bibitem [{\citenamefont {Sheng}, \citenamefont {Mima},\ and\ \citenamefont
		{Zhang}(2005)}]{Sheng2005}%
	\BibitemOpen
	\bibfield  {author} {\bibinfo {author} {\bibfnamefont {Z.~M.}\ \bibnamefont
			{Sheng}}, \bibinfo {author} {\bibfnamefont {K.}~\bibnamefont {Mima}}, \ and\
		\bibinfo {author} {\bibfnamefont {J.}~\bibnamefont {Zhang}},\ }\href
	{\doibase 10.1063/1.2136107} {\bibfield  {journal} {\bibinfo  {journal}
			{Physics of Plasmas}\ }\textbf {\bibinfo {volume} {12}},\ \bibinfo {pages}
		{1} (\bibinfo {year} {2005})}\BibitemShut {NoStop}%
	\bibitem [{\citenamefont {Petrenko}, \citenamefont {Lotov},\ and\ \citenamefont
		{Sosedkin}(2016)}]{Petrenko2016}%
	\BibitemOpen
	\bibfield  {author} {\bibinfo {author} {\bibfnamefont {A.}~\bibnamefont
			{Petrenko}}, \bibinfo {author} {\bibfnamefont {K.}~\bibnamefont {Lotov}}, \
		and\ \bibinfo {author} {\bibfnamefont {A.}~\bibnamefont {Sosedkin}},\ }\href
	{\doibase 10.1016/J.NIMA.2016.01.063} {\bibfield  {journal} {\bibinfo
			{journal} {Nuclear Instruments and Methods in Physics Research Section A:
				Accelerators, Spectrometers, Detectors and Associated Equipment}\ }\textbf
		{\bibinfo {volume} {829}},\ \bibinfo {pages} {63} (\bibinfo {year}
		{2016})}\BibitemShut {NoStop}%
	\bibitem [{\citenamefont {Faure}\ \emph {et~al.}(2010)\citenamefont {Faure},
		\citenamefont {Rechatin}, \citenamefont {Lundh}, \citenamefont {Ammoura},\
		and\ \citenamefont {Malka}}]{Faure2010}%
	\BibitemOpen
	\bibfield  {author} {\bibinfo {author} {\bibfnamefont {J.}~\bibnamefont
			{Faure}}, \bibinfo {author} {\bibfnamefont {C.}~\bibnamefont {Rechatin}},
		\bibinfo {author} {\bibfnamefont {O.}~\bibnamefont {Lundh}}, \bibinfo
		{author} {\bibfnamefont {L.}~\bibnamefont {Ammoura}}, \ and\ \bibinfo
		{author} {\bibfnamefont {V.}~\bibnamefont {Malka}},\ }\href {\doibase
		10.1063/1.3469581} {\bibfield  {journal} {\bibinfo  {journal} {Physics of
				Plasmas}\ }\textbf {\bibinfo {volume} {17}},\ \bibinfo {pages} {083107}
		(\bibinfo {year} {2010})}\BibitemShut {NoStop}%
	\bibitem [{\citenamefont {Lindholm}\ \emph {et~al.}(2008)\citenamefont
		{Lindholm}, \citenamefont {Nickolls}, \citenamefont {Oberman},\ and\
		\citenamefont {Montrym}}]{Lindholm2008}%
	\BibitemOpen
	\bibfield  {author} {\bibinfo {author} {\bibfnamefont {E.}~\bibnamefont
			{Lindholm}}, \bibinfo {author} {\bibfnamefont {J.}~\bibnamefont {Nickolls}},
		\bibinfo {author} {\bibfnamefont {S.}~\bibnamefont {Oberman}}, \ and\
		\bibinfo {author} {\bibfnamefont {J.}~\bibnamefont {Montrym}},\ }in\ \href
	{\doibase 10.1109/MM.2008.31} {\emph {\bibinfo {booktitle} {IEEE Micro}}},\
	Vol.~\bibinfo {volume} {28}\ (\bibinfo {year} {2008})\ pp.\ \bibinfo {pages}
	{39--55}\BibitemShut {NoStop}%
	\bibitem [{\citenamefont {Yee}(1966)}]{Yee1966}%
	\BibitemOpen
	\bibfield  {author} {\bibinfo {author} {\bibfnamefont {K.~S.}\ \bibnamefont
			{Yee}},\ }\href {\doibase 10.1109/TAP.1966.1138693} {\enquote {\bibinfo
			{title} {{Numerical Solution of Initial Boundary Value Problems Involving
					Maxwell's Equations in Isotropic Media}},}\ } (\bibinfo {year}
	{1966})\BibitemShut {NoStop}%
	\bibitem [{\citenamefont {Boris}(1970)}]{Boris1970}%
	\BibitemOpen
	\bibfield  {author} {\bibinfo {author} {\bibfnamefont {J.~P.}\ \bibnamefont
			{Boris}},\ }\href@noop {} {\bibfield  {journal} {\bibinfo  {journal}
			{Proceeding of Fourth Conference on Numerical Simulations of Plasmas}\ }
		(\bibinfo {year} {1970})}\BibitemShut {NoStop}%
	\bibitem [{\citenamefont {Esirkepov}(2001)}]{Esirkepov2001}%
	\BibitemOpen
	\bibfield  {author} {\bibinfo {author} {\bibfnamefont {T.}~\bibnamefont
			{Esirkepov}},\ }\href {\doibase 10.1016/S0010-4655(00)00228-9} {\bibfield
		{journal} {\bibinfo  {journal} {Computer Physics Communications}\ }\textbf
		{\bibinfo {volume} {135}},\ \bibinfo {pages} {144} (\bibinfo {year}
		{2001})}\BibitemShut {NoStop}%
	\bibitem [{\citenamefont {Annenkov}\ \emph
		{et~al.}(2018{\natexlab{b}})\citenamefont {Annenkov}, \citenamefont
		{Berendeev}, \citenamefont {Timofeev},\ and\ \citenamefont
		{Volchok}}]{Annenkov2018}%
	\BibitemOpen
	\bibfield  {author} {\bibinfo {author} {\bibfnamefont {V.~V.}\ \bibnamefont
			{Annenkov}}, \bibinfo {author} {\bibfnamefont {E.~A.}\ \bibnamefont
			{Berendeev}}, \bibinfo {author} {\bibfnamefont {I.~V.}\ \bibnamefont
			{Timofeev}}, \ and\ \bibinfo {author} {\bibfnamefont {E.~P.}\ \bibnamefont
			{Volchok}},\ }\href {\doibase 10.1063/1.5048245} {\bibfield  {journal}
		{\bibinfo  {journal} {Physics of Plasmas}\ }\textbf {\bibinfo {volume}
			{25}},\ \bibinfo {pages} {113110} (\bibinfo {year}
		{2018}{\natexlab{b}})}\BibitemShut {NoStop}%
	\bibitem [{\citenamefont {Astrelin}\ \emph {et~al.}(2016)\citenamefont
		{Astrelin}, \citenamefont {Kandaurov}, \citenamefont {Kurkuchekov},\ and\
		\citenamefont {Trunev}}]{Astrelin2016}%
	\BibitemOpen
	\bibfield  {author} {\bibinfo {author} {\bibfnamefont {V.~T.}\ \bibnamefont
			{Astrelin}}, \bibinfo {author} {\bibfnamefont {I.~V.}\ \bibnamefont
			{Kandaurov}}, \bibinfo {author} {\bibfnamefont {V.~V.}\ \bibnamefont
			{Kurkuchekov}}, \ and\ \bibinfo {author} {\bibfnamefont {Y.~A.}\ \bibnamefont
			{Trunev}},\ }in\ \href {\doibase 10.1063/1.4964175} {\emph {\bibinfo
			{booktitle} {AIP Conference Proceedings}}},\ Vol.\ \bibinfo {volume} {1771}\
	(\bibinfo  {publisher} {AIP Publishing LLC},\ \bibinfo {year} {2016})\ p.\
	\bibinfo {pages} {030019}\BibitemShut {NoStop}%
\end{thebibliography}
\end{document}